\pdfoutput=1
\RequirePackage{ifpdf}
\ifpdf 
\documentclass[pdftex]{sigma}
\else
\documentclass{sigma}
\fi

\begin{document}

\allowdisplaybreaks

\renewcommand{\PaperNumber}{124}

\FirstPageHeading

\renewcommand{\thefootnote}{$\star$}

\ShortArticleName{2D SUSY  QM: Two Fixed Centers}

\ArticleName{Two-Dimensional  Supersymmetric  Quantum \\ Mechanics:
Two Fixed Centers of Force\footnote{This
paper is a contribution to the Proceedings of the Seventh
International Conference ``Symmetry in Nonlinear Mathematical
Physics'' (June 24--30, 2007, Kyiv, Ukraine). The full collection
is available at
\href{http://www.emis.de/journals/SIGMA/symmetry2007.html}{http://www.emis.de/journals/SIGMA/symmetry2007.html}}}

\Author{M.A. GONZ\'ALEZ LE\'ON~$^\dag$, J. MATEOS GUILARTE~$^\ddag$
and M. DE LA TORRE MAYADO~$^\ddag$}

\AuthorNameForHeading{M.A. Gonz\'alez Le\'on, J. Mateos Guilarte
and M. de la Torre Mayado}

\Address{$^\dag$~Departamento de Matem\'atica Aplicada, Universidad de Salamanca, Spain}
\EmailD{\href{mailto:magleon@usal.es}{magleon@usal.es}}

\Address{$^\ddag$~Departamento de F\'{\i}sica Fundamental,
Universidad de Salamanca, Spain}
\EmailD{\href{mailto:guilarte@usal.es}{guilarte@usal.es}, \href{mailto:marina@usal.es}{marina@usal.es}}

\ArticleDates{Received October 09, 2007, in f\/inal form December 10, 2007; Published online December 21, 2007}

\Abstract{The problem of building supersymmetry in the quantum
mechanics of two Coulombian centers of force is analyzed. It is
shown that there are essentially two ways of proceeding. The
spectral problems of the SUSY (scalar) Hamiltonians are quite
similar and become tantamount to solving entangled families of
Razavy and Whittaker--Hill equations in the f\/irst approach. When the
two centers have the same strength, the Whittaker--Hill equations
reduce to Mathieu equations. In the second approach, the spectral
problems are much more dif\/f\/icult to solve but one can still f\/ind the
zero-energy ground states.}


\Classification{81Q60; 81V55}

\section{Introduction}\label{sec1}

Supersymmetry is a bold idea which arose in the seventies in string
and f\/ield theory. It was immediately realized that mechanisms of
spontaneous supersymmetry breaking should be investigated searching
for explanations of the apparent lack of supersymmetry in nature. In
a series of papers \cite{Wit1,Wit2,Wit3} Witten  proposed to analyze this phenomenon in the simplest
possible setting: supersymmetric quantum mechanics. A new area of
research in quantum mechanics was born with far-reaching
consequences both in mathematics and physics.

Of course, there were antecedents in ordinary quantum mechanics
({\it nihil novum sub sole}), and indeed even before. The track can
be followed back to some work by Clif\/ford on the Laplacian operator,
see~\cite{Cliff}, quoted in~\cite{Hit}. Recast in modern SUSY
language, the Clif\/ford supercharge is:
\[
Q=\left(\begin{array}{lc} 0 & i\nabla_1+j\nabla_2+k\nabla_3
\\ 0 & 0 \end{array}\right), \qquad i^2=j^2=k^2=-1,
\qquad ij=-ji=k \,\, \, {\rm ciclyc},
\]
where $i$, $j$, $k$ are the imaginary unit quaternions and
\[
\nabla_1=\frac{\partial}{\partial x_1}+A_1(\vec{x}), \qquad
\nabla_2=\frac{\partial}{\partial x_2}+A_2(\vec{x}), \qquad
\nabla_3=\frac{\partial}{\partial x_3}+A_3(\vec{x})
\]
are the components of the gradient modif\/ied by the components of the
electromagnetic vector potential. The SUSY Hamiltonian is:
\[
Q^\dagger Q+QQ^\dagger=\left(\!\begin{array}{l@{}c} -\bigtriangleup +i
B_1(\vec{x})+j B_2(\vec{x})+ k B_3(\vec{x}) & 0
\\ 0 & -\bigtriangleup -i
B_1(\vec{x})-j B_2(\vec{x})- k B_3(\vec{x})\end{array}\!\right)\! ,
\]
the Laplacian plus Pauli terms. Needless to say, an identical
construction relates the Dirac operator in electromagnetic and/or
gravitational f\/ields backgrounds with the Klein--Gordon operator. The
factorization method of identifying the spectra of Schrodinger
operators, see~\cite{Inf} for a review, is another antecedent of
supersymmetric quantum mechanics that can also be traced back to the
19th century through the Darboux theorem.

In its modern version, supersymmetric quantum mechanics prompted the
study of many one-dimensional systems from a physical point of view.
A good deal of this work can be found in~\cite{Casal,
Ber,Khare,Junker}. A frequent starting point
in this framework is the following problem: given a non-SUSY
one-dimensional quantum Hamiltonian, is it possible to build a
supersymmetric extension? The answer to this question is positive
when one f\/inds a solution to the Riccati equation
\[
{1\over 2}\frac{dW}{dx} \frac{dW}{dx}+{\hbar\over
2}\frac{d^2W}{dx^2}=V(x),
\]
that identif\/ies the~-- a priori unknown~-- ``superpotential'' $W(x)$
from the~--~given~-- poten\-tial~$V(x)$. Several examples of this strategy
have been worked out in~\cite{Car}.

The formalism of physical supersymmetric systems with more than one
degree of freedom was f\/irst developed by Andrianov, Iof\/fe and
coworkers in a series of papers~\cite{Ioffe,Ioffe2},
published in the eighties. The same authors, almost simultaneously,
considered higher than one-dimensional SUSY quantum mechanics from
the point of view of the factorization of $N$-dimensional quantum
systems~\cite{Andr,Andr1}. Factorability, even though
essential in $N$-dimensional SUSY quantum mechanics, is not so
ef\/fective as compared with the one-dimensional situation. Some
degree of separability is also necessary to achieve analytical
results. For this reason we started a program of research in the
two-dimensional supersymmetric classical mechanics of Liouville
systems~\cite{Perelomov}; i.e., those separable in elliptic, polar,
parabolic, or Cartesian coordinates, see the papers \cite{AM} and
\cite{AoP}. We followed this path in the quantum domain for Type I
Liouville models in \cite{JPA}, whereas Iof\/fe et al.\ also studied
the interplay between supersymmetry and integrability in quantum and
classical settings in other type of models in~\cite{Andr2,Can}. In these papers, a new structure was
introduced: second-order supercharges provided intertwined scalar
Hamiltonians even in the two-dimensional case, see~\cite{Ioffe3} for
a review. This higher-order SUSY algebra allows for new forms of
non-conventional separability in two dimensions. There are two
possibilities: (1)~a~similarity transformation performs separation
of variables in the supercharges and some eigenfunctions (partial
solvability) can be found, see~\cite{Ioffe4,Ioffe5}. (2)~One
of the two intertwined Hamiltonian allows for exact separability:
the spectrum of the other is consequently known~\cite{Ioffe6,Ioffe7}.

The second dif\/f\/iculty with the jump in dimensions is the
identif\/ication of the superpotential. Instead of the Riccati
equation one must solve the PDE:
\[
{1\over
2}\vec{\nabla}W(\vec{x})\cdot\vec{\nabla}W(\vec{x})+{\hbar\over
2}\nabla^2W(\vec{x})=V(\vec{x})  .
\]
In our case, we look for solutions of this PDE when $V(\vec{x})$ is
the potential energy of the two Coulombian centers. We do not know
how to solve it in general, but two dif\/ferent strategies should help
us. First, following the work in \cite{KLPW} and \cite{KLPW1} on the
supersymmetric Coulomb problem, we shall choose the superpotential
as the solution of the Poisson equation:
\[
{\hbar\over 2}\nabla^2W(\vec{x})=V(\vec{x})  .
\]
The superpotential will be the solution of {\it another}
Riccati-like PDE where a classical piece must be added to the
potential of the two centers. Second, as in~\cite{Manton,Heumann} the selection of superpotential requires the solution
of the Hamilton--Jacobi equation:
\[
{1\over
2}\vec{\nabla}W(\vec{x})\cdot\vec{\nabla}W(\vec{x})=V(\vec{x})
.
\]
Again the superpotential must solve a third Riccati-like PDE, where
now a quantum piece must be added to the potential of the two
centers.

The organization of the paper is as follows: We start by brief\/ly
recalling the non-SUSY classical, Newtonian~\cite{landau}, and
quantal, Coulombian~\cite{Pauling,Greiner}, two-center
problem. We shall constrain the particle to move in one plane
containing the two centers. The third coordinate is cyclic and it
would be easy to extend our results to three dimensions. In Section~\ref{sec2} the formalism of two-dimensional SUSY quantum mechanics is
developed, and the superpotential of the f\/irst Type is identif\/ied
for two Coulombian centers. Bosonic zero-energy ground states are
also found. Sections~\ref{sec3} and~\ref{sec4} are devoted to formulating the SUSY
system in elliptic coordinates where the problem is separable in
order to f\/ind fermionic zero-energy ground states. It is also shown
that the spectral problem is tantamount to families of two ODE's of
Razavy~\cite{Razavy1}, and Whittaker--Hill type~\cite{Razavy2}, see
also~\cite{Bondar}. Since these systems are quasi-exactly solvable,
several eigenvalues are found following the work in~\cite{FGR}. In
Section~\ref{sec5} two centers of the same strength are studied and some
eigenfunctions are also found. Section~\ref{sec6} is fully devoted to the
analysis of the Manton--Heumann approach applied to the two
Coulombian centers. Finally, a summary is of\/fered in Section~\ref{sec7}.

\subsection{The classical problem of two Newtonian/Coulombian
centers}\label{sec1.1}

\begin{figure}[h]\centering
\includegraphics[height=3.5cm]{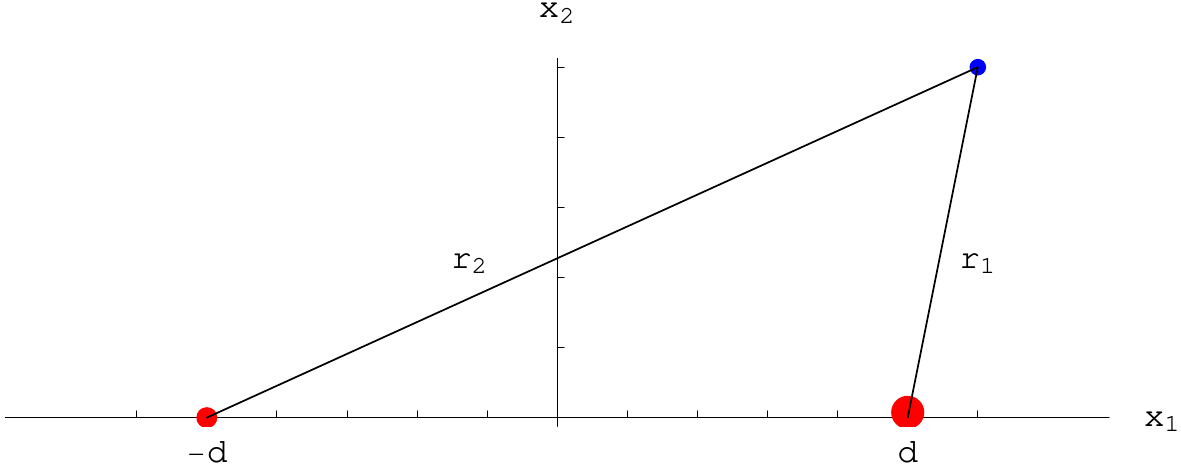}

\caption{Location of the two centers and distances
to the particle from the centers.}
\end{figure}

The classical action for a system of a light particle moving in a
plane around two heavy bodies which are sources of static
Newtonian/Coulombian forces is:
\[ \tilde{S} =\int d t \left\{ {1\over 2}\, m \left( {d x_1 \over d
t}{d x_1 \over d t} + {d x_2 \over d t}{d x_2 \over d t} \right) -
{\alpha_1\over r_1} - {\alpha_2 \over r_2} \right\}.
\]
The centers are located at the points $(x_1=-d$, $x_2=0)$,
$(x_1=d$, $x_2=0)$, their strengths are $\alpha_1=\alpha \geq
\alpha_2=\delta\alpha
> 0 $, $\delta \in (0,1]$, and
\[
r_1=\sqrt{(x_1-d)^2 + x_2^2} , \qquad
r_2=\sqrt{(x_1+d)^2+x_2^2}
\]
are the distances from the particle to the centers. In the following
formulas we show the dimensions of the coupling constants and
parameters and def\/ine non-dimensional variables:
\begin{gather*}
[\alpha_1] = [\alpha_2]=[\alpha] = M L^3 T^{-2}, \qquad
[d]=L, \qquad [\delta]=1,\\
x_1\rightarrow  d\, x_1, \qquad  x_2\rightarrow  d\, x_2
, \qquad  t\rightarrow \sqrt{\frac{d^3m}{\alpha}}\, t,
\\
r_1\rightarrow d\, r_1=d\sqrt{(x_1-1)^2 + x_2^2}, \qquad
r_2\rightarrow d\, r_2=d\sqrt{(x_1+1)^2+x_2^2}.
\end{gather*}
In the rest of the paper we shall use non-dimensional variables.
From the non-dimensional action
\[ \tilde{S} = \sqrt{md\alpha}\,
S=\sqrt{md\alpha} \int \, dt \, \left\{ {1\over 2} \left( {d x_1
\over dt}{d x_1 \over dt} + {d x_2 \over dt}{d x_2 \over dt} \right)
- {1\over r_1} - {\delta \over r_2} \right\},
\]
the linear momenta and Hamiltonian are def\/ined:
\begin{gather*}
p_1=\frac{\partial L}{\partial \dot{x_1}}=\frac{dx_1}{dt} , \qquad  p_2=\frac{\partial L}{\partial
\dot{x_2}}=\frac{dx_2}{dt},
\\
\tilde{H}=\frac{\alpha}{d} H  ,  \qquad  H=
{1\over 2} (p_1^2+ p_2^2) +{1\over r_1} + {\delta \over r_2} .
\end{gather*}
This system is completely integrable because there  exists a ``second invariant'' in involution with the Hamiltonian:
\[ \tilde{I}_2
= (m d \alpha) I_2 , \qquad I_2={1\over 2} (l^2 - p_2^2)+x_1
\left( {\delta \over r_2} -{1\over r_1} \right), \qquad l^2 =
(x_1 p_2-x_2 p_1)^2.
\]

\subsection{The quantum problem of two Coulombian centers of force}\label{sec1.2}

 If $\sqrt{m\alpha d}$ is of the order of the Planck constant
 $\hbar$,
the system is of quantum nature. Canonical quantization in terms of
the non-dimensional $\bar\hbar$ constant,
\begin{gather*}
p_i \rightarrow \hat{p}_i=-i\bar{\hbar}\frac{\partial}{\partial x_i}
 , \qquad x_i \rightarrow \hat{x}_i=x_i,
\\
[\hat{x}_i,\hat{p_j}]=i\bar{\hbar}\delta_{ij} ,
\qquad \bar{\hbar}=\frac{\hbar}{\sqrt{md\alpha }},
\end{gather*}
converts the dynamical variables into operators. The quantum
Hamiltonian, $\hat{\tilde{H}}=\frac{\alpha}{d}\hat{H}$, and the
quantum symmetry operator, $\hat{\tilde{I}}_2=(m d \alpha)
\hat{I}_2$, are mutually commuting operators:
\begin{gather*} \hat{H}=
-{\bar{\hbar}\over 2} \left( \frac{\partial^2}{\partial x_1^2}+
\frac{\partial^2}{\partial x_2^2}\right) + {1\over r_1} + {\delta
\over r_2}, \qquad [\hat{H},
\hat{I}_2]=\hat{H}\hat{I}_2-\hat{I}_2\hat{H}=0 ,
\\ \hat{I}_2= -{\bar{\hbar}^2\over 2} \left( (x_1^2-1)
\frac{\partial^2}{\partial x_2^2} + x_2^2 \frac{\partial^2}{\partial
x_1^2} - 2 x_1 x_2 \frac{\partial^2}{\partial x_1 \partial x_2} -
x_1 \frac{\partial}{\partial x_1} - x_2 \frac{\partial}{\partial
x_2} \right) + x_1 \left( {\delta \over r_2} - {1 \over r_1} \right)
 .
\end{gather*}

\section[Two-dimensional ${\cal N}=2$ SUSY quantum mechanics]{Two-dimensional $\boldsymbol{{\cal N}=2}$ SUSY quantum mechanics}\label{sec2}

We now describe how to build a non-specif\/ic system in
two-dimensional ${\cal N}=2$ SUSY quantum mechanics. Besides
commuting -- non-commuting -- operators there are anti-commuting~-- non-anti-commuting operators to be referred respectively as
``bosonic'' and ``fermionic'' by analogy with QFT. The
Fermi operators are represented on Euclidean spinors in ${\mathbb
R}^4$ by the Hermitian $4\times 4$ gamma matrices:
\begin{gather*} \psi^j_1=\frac{i}{\sqrt{2}}\gamma^j ,
\qquad \psi^j_2=-\frac{i}{\sqrt{2}}\gamma^{2+j}, \qquad
(\gamma^j)^\dagger=\gamma^j , \qquad (\gamma^{2+j})^\dagger=\gamma^{2+j},\\
\{\gamma^j,\gamma^k\}=2\delta^{jk}=\{\gamma^{2+j},\gamma^{2+k}\}
, \qquad \{\gamma^j,\gamma^{2+k}\}=0, \qquad
j,k=1,2.
\end{gather*}

The building blocks of the SUSY system are the two (${\cal N}=2$)
quantum Hermitian  supercharges: $\hat{Q}_1^\dagger=\hat{Q}_1$,  $\hat{Q}_2^\dagger=\hat{Q}_2$,
\[ \hat{Q}_1=\sqrt{{\bar{\hbar}}} \sum_{j=1}^2
\left(-i{\bar{\hbar}}{\partial\over\partial
x_j} \psi_1^j-\frac{\partial W}{\partial
x_j} \psi_2^j\right), \qquad
\hat{Q}_2=\sqrt{{\bar{\hbar}}} \sum_{j=1}^2
\left(-i{\bar{\hbar}}{\partial\over\partial
x_j} \psi_2^j+\frac{\partial W}{\partial
x_j} \psi_1^j\right).
\]
It is convenient to def\/ine the non-Hermitian supercharges
$\hat{Q}_\pm=\hat{Q}_1\pm i\hat{Q}_2$,
\begin{gather*}
\hat{Q}_+=i\sqrt{\bar{\hbar}} \left(\begin{array}{cccc}
0 & 0 & 0 & 0 \vspace{1mm}\\
\bar{\hbar}{\partial \over \partial x_1}-{\partial W \over
\partial x_1} & 0 & 0 & 0 \vspace{1mm}\\ \bar{\hbar}{\partial \over\partial
x_2}-{\partial W \over
\partial x_2} & 0 & 0 & 0
\vspace{1mm}\\ 0 & -\bar{\hbar}{\partial \over\partial x_2}+{\partial W \over\partial x_2}
& \bar{\hbar}{\partial \over\partial x_1}-{\partial W \over\partial
x_1} & 0
\end{array}\right) ,
\\
\hat{Q}_-=i\sqrt{\bar{\hbar}} \left(\begin{array}{cccc} 0 &
 \bar{\hbar}{\partial \over \partial
x_1}+{\partial W \over \partial x_1} & \bar{\hbar} {\partial \over
\partial x_2}+{\partial W \over \partial x_2} & 0
\vspace{1mm}\\ 0 & 0 & 0 & -\bar{\hbar}{\partial \over \partial x_2}-{\partial W \over \partial x_2}
\vspace{1mm}\\ 0 & 0 & 0 & \bar{\hbar}{\partial \over \partial x_1}+{\partial W \over \partial x_1}
\vspace{1mm}\\ 0 & 0 & 0 & 0
\end{array}\right)
\end{gather*}
because their anti-commutator determines the Hamiltonian $\hat{H}_S$
of the supersymmetric system:
\[ \{\hat{Q}_+ , \hat{Q}_- \}=2 \bar{\hbar} \hat{H}_S,
\qquad [\hat{Q}_+,\hat{H}_S]=[\hat{Q}_-,\hat{H}_S]=0.
\]
The explicit form of the quantum SUSY Hamiltonian is enlightened by
the ``Fermi'' number $F= \sum\limits_{j=1}^2
\psi_+^j\psi_-^j$ operator:
\[
\hat{H}_S=\left(\begin{array}{cccc} \hat{h}^{(0)} & 0 & 0 &0 \\
0 & \hat{h}^{(1)}_{11} & \hat{h}^{(1)}_{12} & 0\\ 0 &
\hat{h}^{(1)}_{21}& \hat{h}^{(1)}_{22} & 0
\\ 0 & 0 & 0 & \hat{h}^{(2)}
\end{array}\right) , \qquad F={\displaystyle \sum_{j=1}^2}
\psi_+^j\psi_-^j=\left(\begin{array}{cccc} 0 & 0 & 0 & 0
\\ 0 & 1 & 0 & 0 \\ 0 & 0 & 1 & 0 \\ 0 & 0 & 0 &
2\end{array}\right).
\]
It has a block diagonal structure acting on the sub-spaces of the
Hilbert space of Fermi numbers 0, 1, and 2. In the sub-spaces of
Fermi numbers even ${\hat H}_S$ acts as ordinary dif\/ferential
Schr\"odinger operators. The scalar Hamiltonians are:
\begin{gather*}
2\hat{h}^{(f=0)}=-\bar{\hbar}^2\nabla^2+\vec{\nabla}W\vec{\nabla}W+\bar{\hbar}
\nabla^2W = -\bar{\hbar}^2\nabla^2 + 2 {\hat V}^{(0)},
\\  2\hat{h}^{(f=2)}=
-\bar{\hbar}^2\nabla^2+\vec{\nabla}W\vec{\nabla}W-\bar{\hbar}
\nabla^2W =-\bar{\hbar}^2\nabla^2 + 2 {\hat V}^{(2)}.
\end{gather*}
In the sub-space of Fermi number 1, however, ${\hat H}_S$ is a
matrix of dif\/ferential operators, the $2\times 2$ matrix
Hamiltonian:
\begin{gather*} 2\hat{h}^{(f=1)}=\left(\begin{array}{cc}
-\bar{\hbar}^2\nabla^2+\vec{\nabla}W\vec{\nabla}W-\bar{\hbar}
\Box^2W &
-2\bar{\hbar} {\partial^2 W \over\partial x_1\partial x_2} \vspace{2mm}\\
-2\bar{\hbar}{\partial^2 W \over\partial x_1\partial x_2} &
-\bar{\hbar}^2\nabla^2+\vec{\nabla}W\vec{\nabla}W+\bar{\hbar}\Box^2W
\end{array}\right) ,
\\ \vec{\nabla}=\frac{\partial}{\partial x_1}\cdot
\vec{e}_1+\frac{\partial}{\partial x_2}\cdot \vec{e}_2 , \qquad
\nabla^2={\partial^2\over\partial x_1\partial
x_1}+{\partial^2\over\partial x_2\partial x_2} , \qquad
   \Box^2={\partial^2\over\partial x_1\partial
x_1}-{\partial^2\over\partial x_2\partial x_2}.
\end{gather*}
This is exactly the structure unveiled in \cite{Ioffe,Ioffe2}. All the interactions expressed in ${\hat H}_S$ come
from the as yet unspecif\/ied function $W(x_1,x_2)$, which is thus
called the superpotential.

\subsection{The superpotential I for the two-center
problem}\label{sec2.1}

To build a supersymmetric system containing the interactions due to
two Coulombian centers of force, we must start by identifying the
superpotential. One possible choice, inspired by \cite{KLPW}, is
having the two-center potential energy in $\hat{h}^{(0)}$ in the
term proportional to $\bar\hbar$. We must therefore solve the
Poisson equation to f\/ind the superpotential I:
\begin{equation} {\bar{\hbar} \over 2} \nabla^2 \hat{W} = -{1\over
r_1} -{\delta\over r_2} , \qquad \hat{W}(x_1,x_2)=-\frac{2
r_1}{\bar{\hbar}} - \frac{2 \delta r_2}{\bar{\hbar}} .
\label{eq:sups}
\end{equation}
Note that the anticommutator between the supercharges induces a
$\bar{\hbar}$ factor in front of the Laplacian. This fact, in turns,
forces the singularity of the superpotential (henceforth, also of
the potential) at the classical limit $\bar{\hbar}=0$. The same
singularity arises in the bound state spectra of atoms, e.g., in the
energy levels of the hydrogen atom. The potential energies in the
scalar sectors are accordingly:
\[
{\hat V}_I^{(0)\choose (2)}=\frac{2}{\bar{\hbar}^2 } \left[ 1 +
\delta^2 + \delta \left( {r_1\over r_2} +{r_2\over r_1} -{4 \over
r_1 r_2} \right) \right] \mp  \left({ 1\over r_1} +{\delta \over
r_2} \right).
\]

\begin{figure}[h]
\centering \includegraphics[height=4.5cm]{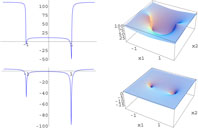}
\includegraphics[height=4.5cm]{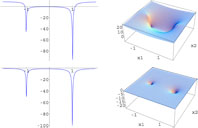}

\caption{Cross section ($x_2=0$) and 3D
graphics of the quantum potential $\hat{V}^{(0)}$ for $\delta=1/2$.
Cases: Upper row: (a) $\bar{\hbar}=0.2$, (b) $\bar{\hbar}=0.4$.
Lower row: (a) $\bar{\hbar}=1$ and (b) $\bar{\hbar}=10$. Increasing
$\bar{\hbar}$ the centers become more and more attractive.}
\end{figure}

We stress that the superpotential I is a solution of the
Riccati-like PDE's:
\[ \vec{\nabla}\hat{W}\vec{\nabla}\hat{W}\pm\bar{\hbar}
\nabla^2{\hat W} =2{\hat V}_I^{(0)\choose (2)}
\]
and the scalar Hamiltonians for two SUSY Coulombian centers read:
\[ \hat{h}^{(0)\choose(2)}=-{\bar{\hbar}^2\over 2}\nabla^2+
\frac{2}{\bar{\hbar}^2 } \left[ 1 + \delta^2 + \delta \left(
{r_1\over r_2} +{r_2\over r_1} -{4 \over r_1 r_2} \right) \right]\mp
 \left({ 1\over r_1} +{\delta \over r_2} \right) .
\]

\subsection{Bosonic zero modes I}\label{sec2.2}

The bosonic zero modes
\[ \hat{Q}_\pm\Psi_0^{(0)}(x_1,x_2)=0 , \qquad
\hat{Q}_\mp\Psi_0^{(2)}(x_1,x_2)=0 ,
\]
if normalizable, are the bosonic ground states of the system:
\[
\Psi_0^{(0)}(x_1,x_2)=  \left(
\begin{array}{c} {\rm exp}[{(-2 r_1-2 \delta r_2) \over \bar{\hbar}^2}]
\\ 0 \\ 0 \\ 0
\end{array} \right), \qquad
 \Psi_0^{(2)}(x_1,x_2)= \left(
\begin{array}{c} 0
\\ 0 \\ 0 \\ {\rm exp}[{(2 r_1+2 \delta r_2) \over
\bar{\hbar}^2}]\end{array}\right) .
\]

The norm of the true bosonic ground state $\Psi_0^{(0)}$ is f\/inite
and given in terms of Bessel functions:
\begin{gather*}
N(\bar{\hbar})=\int_{-\infty}^\infty  \int_{-\infty}^\infty dx_1dx_2
\, e^{\frac{2}{\bar{\hbar}} \hat{W}(x_1,x_2)}= 2 \int_{0}^\infty
dx_1 \int_{0}^\infty dx_2 \, e^{\frac{2}{\bar{\hbar}^2} (-2
r_1-2\delta r_2)},\\
 N(\bar{\hbar})=2\pi \left[
\frac{\bar{\hbar}^2}{4(1+\delta)}
I_0\left(\frac{4}{\bar{\hbar}^2}(1-\delta)\right)
K_1\left(\frac{4}{\bar{\hbar}^2}(1+\delta)\right)\right.\\
\left. \phantom{N(\bar{\hbar})=}{} +\frac{\bar{\hbar}^2}{4(1-\delta)}
I_1\left(\frac{4}{\bar{\hbar}^2}(1-\delta)\right)
K_0\left(\frac{4}{\bar{\hbar}^2}(1+\delta)\right)\right] .
\end{gather*}

In Fig.~3 plots of the zero energy bosonic ground state
probability density of f\/inding the particle in some area of the
plane are shown for several values of $\bar\hbar$. The drawings
reveal the physical meaning of the $\bar\hbar=0$ singularity: for
$\bar\hbar=0.2$ we see the particle probability density peaked
around the center on the right with a very small probability.
Exactly at $\bar\hbar=0$, $e^{-{4 r_1\over\bar\hbar^2}}$ is f\/inite
only for $r_1=0$ and zero otherwise, giving probability of f\/inding
the particle exactly in the center. $e^{-{4 \delta
r_2\over\bar\hbar^2}}$, however, is zero $\forall\, r_2$ meaning that
the probability of this state is zero at the classical limit; in
classical mechanics there are no isolated discrete states.

\begin{figure}[h]\centering
\includegraphics[height=2.5cm]{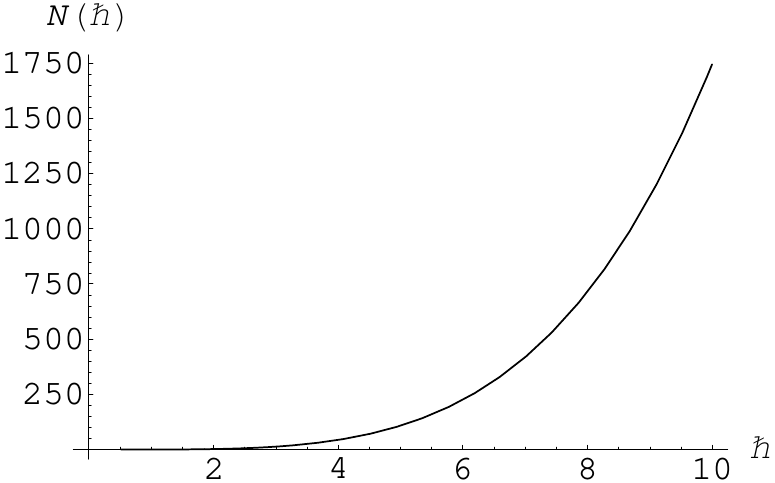}
\includegraphics[height=3.cm]{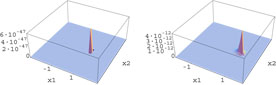}

\includegraphics[height=3.cm]{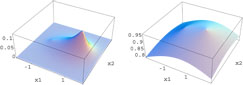}

\caption{Graphics of the norm as a
function of $\bar\hbar$ and the ground state probability density,
$|\Psi_0^{(0)}(x_1,x_2)|^2$, for $\delta=1/2$ and the values of
$\bar{\hbar} = 0.2, \, 0.4,\, 1$ and $10$. Note the extreme smallness
for $\bar{\hbar}=0.2$. The norms for these four cases are:
$N(0.2)= 3.5806 \cdot 10^{-47}$, $N(0.4)= 2.0576 \cdot
10^{-13}$, $N(1)=0.00942$ and $N(10)=1743.94$.}
\end{figure}

\section[Two-dimensional ${\cal N}=2$ SUSY quantum mechanics in elliptic coordinates]{Two-dimensional $\boldsymbol{{\cal N}=2}$ SUSY quantum mechanics\\ in elliptic coordinates}\label{sec3}

The search for more eigenfunctions of the SUSY Hamiltonian requires
the use of the separability in elliptic coordinates of the problem
at hand. This, in turn, needs the translation of our two-dimensional
${\cal N}=2$ SUSY system to elliptic coordinates. A general
reference (in Russian) where SUSY quantum mechanics is formulated in
curvilinear coordinates is \cite{Ioffe8}, see also \cite{Art} to
f\/ind a~more geometric version of SUSY QM on Riemannian manifolds.

The change from Cartesian to elliptic coordinates,
\begin{gather*} x_1=u v \in (-\infty,+\infty) ,\qquad
 x_2=\pm \sqrt{(u^2-1) (1-v^2)} \in (-\infty,+\infty),
\\
u={1\over 2}(r_1+r_2)\in (1,+\infty) , \qquad
v={1\over 2} (r_2-r_1)\in (-1,1),
\end{gather*}
induces a map from the plane to the inf\/inite elliptic strip: $
{\mathbb R}^2\equiv (-\infty,+\infty)\times (-\infty,+\infty)
\Longrightarrow  {\mathbb E}^2\equiv (-1,1)\times (1,+\infty)$.
This map also induces a non-Euclidean (but f\/lat) metric in ${\mathbb
E}^2$:
\[ g(u,v)=\left(
\begin{array}{cc} g_{uu}=\frac{u^2-v^2}{u^2-1} &
g_{uv}=0 \vspace{1mm}\\ g_{vu}=0 &
g_{vv}=\frac{u^2-v^2}{1-v^2}\end{array} \right) ,
\]
with Christof\/fel symbols:
\begin{gather*}
\Gamma_{uu}^u=\displaystyle\frac{-u (1-v^2)}{(u^2-v^2)(u^2-1)}  ,\qquad
\Gamma_{vv}^v=\displaystyle\frac{v (u^2-1)}{(u^2-v^2)(1-v^2)} ,\qquad
\Gamma_{uv}^u=\Gamma_{vu}^u=\displaystyle\frac{-v}{u^2-v^2},
\\ \Gamma_{uu}^v=\displaystyle\frac{v
(1-v^2)}{(u^2-v^2)(u^2-1)}  ,\qquad \Gamma_{vv}^u=\displaystyle\frac{-u
(u^2-1)}{(u^2-v^2)(1-v^2)} ,\qquad
\Gamma_{uv}^v=\Gamma_{vu}^v=\displaystyle\frac{u }{u^2-v^2} .
\end{gather*}
Using the zweig-bein chosen in this form,
\begin{gather*}
g^{uu}(u,v)=\sum_{j=1}^2e^u_j(u,v)e^u_j(u,v) , \qquad
g^{vv}(u,v)=\sum_{j=1}^2e^v_j(u,v)e^v_j(u,v) ,
\\
e^u_1(u,v)=\left(u^2-1\over u^2-v^2\right)^{{1\over 2}},
\qquad e^v_2(u,v)=\left(1-v^2\over u^2-v^2\right)^{{1\over 2}}
\end{gather*}
we now def\/ine ``elliptic'' spinors, ``elliptic'' Fermi
operators, and ``elliptic'' supercharges:
\begin{gather*}
\psi_{\pm}^{u}(u,v)=e^u_1(u,v)\psi^1_\pm , \qquad
\psi_\pm^v(u,v)=e^v_2(u,v)\psi_\pm^2 ,
\\
 \hat{C}_+=-
i \sqrt{\bar{\hbar}} \left(\!\!\begin{array}{cccc} 0 & 0 & 0 & 0 \\
e^u_1\nabla_u^- & 0 & 0 & 0
\\ e^v_2\nabla_v^- & 0 & 0 & 0
\\ 0 & -e^v_2 \left( \nabla_v^- -{\bar\hbar v\over u^2-v^2} \right)
& e^u_1 \left( \nabla_u^- +{\bar\hbar u\over u^2-v^2} \right)  & 0
\end{array}\!\!\right) \!, \qquad \nabla_u^\mp= \bar{\hbar} {\partial\over\partial u}\mp
{d\hat{F}\over du},
\\
\hat{C}_-=-i \sqrt{\bar{\hbar}}  \left(\!\!\begin{array}{cccc} 0 & e^u_1
\left( \nabla_u^+ +{\bar\hbar u\over u^2-v^2} \right) & e^v_2 \left(
\nabla_v^+ -{\bar\hbar v\over u^2-v^2} \right) & 0
\\ 0 & 0 & 0 & -e^v_2\nabla_v^+
\\ 0 & 0 & 0 & e^u_1\nabla_u^+
\\ 0 & 0 & 0 & 0
\end{array}\!\!\right)\! , \qquad \nabla_v^\mp = \bar{\hbar}{\partial\over\partial v}\mp {d\hat{G}\over
dv} ,
\end{gather*}
where  $ \hat{W}(u,v)=\hat{F}(u)+\hat{G}(v)$.

To obtain the supercharges in Cartesian coordinates from the
supercharges in elliptic coordinates, besides expressing $u$ and $v$
in terms of $x_1$ and $x_2$, one needs to act by conjugation with
the idempotent, Hermitian matrix:
\[
{\cal S} = \left( \begin{array}{cccc} 1 & 0 & 0 & 0 \\
0 & - v e^u_1(u,v) & - u e^v_2(u,v) & 0 \\ 0 & - u e^v_2(u,v) & v
e^u_1(u,v) & 0 \\ 0 & 0 & 0 & -1 \end{array} \right), \qquad
{\cal S}  \hat{C}_+ {\cal S} = \hat{Q}_+   , \qquad  {\cal S}
\hat{C}_- {\cal S} = \hat{Q}_- .
\]

 Equation (\ref{eq:sups}) in elliptic coordinates,
\begin{gather} {\bar{\hbar} \over 2} \left[ {u^2 -1 \over
u^2-v^2} \left( {d^2 \hat{F} \over d u^2} + {u\over u^2-1 }{d
\hat{F} \over d u} \right) + {1 -v^2 \over u^2-v^2} \left( {d^2
\hat{G} \over d v^2}-{v\over 1-v^2} {d \hat{G} \over d v} \right)
\right]\nonumber\\
\qquad{} = -{(1+\delta) u \over u^2-v^2} + {(\delta-1) v\over
u^2-v^2} \label{eq:poisse}
\end{gather}
is separable:
\[ (u^2-1) {d^2 \hat{F} \over d u^2} + u {d \hat{F}
\over d u} + \frac{2 (1+\delta) u}{\bar{\hbar}} =\kappa =- (1-v^2)
{d^2 \hat{G}\over d v^2} + v {d \hat{G} \over d v} + \frac{2 (\delta
-1) v}{\bar{\hbar}} ,
\]
with separation constant $\kappa$. The general solution of equation
(\ref{eq:poisse}) depends on two integration constants (besides an
unimportant additive constant):
\begin{gather*}
\hat{W}(u,v;\kappa, C_1,C_2)= - 2 \frac{(1+\delta)}{\bar{\hbar}} u
+ {C_1\over \bar{\hbar}} \ln \big(u+\sqrt{u^2-1}\big) +{\kappa \over
2 \bar{\hbar}} \left( \ln \big(u+\sqrt{u^2-1}\big) \right)^2 \\
\phantom{\hat{W}(u,v;\kappa, C_1,C_2)=}{} + 2
\frac{(1-\delta)}{\bar{\hbar}} v+ {C_2 \over \bar{\hbar}} \arcsin  v -{\kappa \over 2 \bar{\hbar}} \left( \arcsin v
\right)^2.
\end{gather*}
We f\/ind thus a three-parametric family of supersymmetric models for
which the potentials in the scalar sectors are:
\begin{gather*} {\hat V}_I^{(0)\choose
(2)}(x_1, x_2; \kappa, C_1,C_2)=\frac{2}{\bar{\hbar}^2 } \left[ 1
+ \delta^2 + \delta \left( {r_1\over r_2} +{r_2\over r_1} -{4 \over
r_1 r_2} \right) \right]  \\
\qquad{} +\frac{1}{2 \bar{\hbar}^2 r_1 r_2} \Bigg[ \left( C_1+ \kappa \ln
\frac{r_1+r_2+\sqrt{(r_1 + r_2)^2 -  4}}{2}\right)^2 + \left( C_2 -
\kappa \arcsin \frac{r_2-r_1}{2}\right)^2  \\
\qquad{} - 2 (1 +
\delta) \sqrt{(r_1 + r_2)^2- 4 } \left( C_1 + \kappa \ln
\frac{r_1 + r_2 + \sqrt{(r_1 + r_2)^2 - 4}}{2}\right)\\
 \qquad{}+ 2(1-\delta) \sqrt{4-(r_2-r_1)^2} \left( C_2 - \kappa
\arcsin \frac{r_2-r_1}{2 }\right) \Bigg] \mp  \left({ 1\over r_1}
+{\delta \over r_2} \right) .
\end{gather*}
We shall restrict ourselves in the sequel (as before) to the
simplest choice $\kappa=C_1=C_2=0$ such that we shall work with the
``elliptic'' superpotential I
\[
\hat{W}(u,v) = - \frac{2 (1+\delta) u}{\bar{\hbar}} +\frac{ 2
(1-\delta) v}{\bar{\hbar}} , \qquad \hat{W}(x_1,x_2)=-\frac{2
r_1}{\bar{\hbar}} - \frac{2 \delta r_2}{\bar{\hbar}} ,
\]
because this election is signif\/icative and contains enough
complexity.

Considering families of superpotentials related to the same physical
system in our 2D framework dif\/fers from a similar analysis on the 1D
SUSY oscillator, see e.g.~\cite{Mielnik}, in two aspects: (a)~Because we solve the Poisson equation, not the Riccati equation, the
family of superpotentials induces dif\/ferent families of potentials in
both $\hat{h}^{(0)}$ and $\hat{h}^{(2)}$. (b) $\hat{h}^{(0)}$ and
$\hat{h}^{(2)}$ are not iso-spectral because are not directly
intertwined. The spectrum of $\hat{h}^{(1)}$ is the union of the
spectra of $\hat{h}^{(0)}$ and $\hat{h}^{(2)}$.

\subsection{Fermionic zero modes I}\label{sec3.1}

The fermionic zero modes
\begin{gather*} \hat{C}_+ \Psi_0^{(1)}(u,v)=0 \quad , \quad
\hat{C}_-\Psi_0^{(1)}(u,v)=0 , \qquad
\Psi_0^{(1)}(u,v)= \left(\begin{array}{c} 0 \\
\psi^{(1)1}_0(u,v) \\ \psi^{(1)2}_0(u,v) \\ 0
\end{array}\right),
\\
\Psi_0^{(1)}(u,v)= \left(\begin{array}{c} 0 \\
\psi^{(1)1}_0(u,v) \\ \psi^{(1)2}_0(u,v) \\ 0
\end{array}\right)=\frac{A_1}{\sqrt{u^2-v^2}}\left(
\begin{array}{c} 0\\
e^{\frac{-\hat{F}(u)+\hat{G}(v)}{\bar{\hbar}}}\\ 0 \\ 0
\end{array} \right)\ +\  \frac{A_2}{\sqrt{u^2-v^2}}\left(
\begin{array}{c} 0\\0\\
e^{\frac{\hat{F}(u)-\hat{G}(v)}{\bar{\hbar}}}\\ 0
\end{array} \right)
\end{gather*}
are fermionic ground states if normalizable. Because the norm is:
\begin{gather*} N(\bar{\hbar})=2\int_{-1}^1 \!\! dv\int_{1}^{\infty} \!\! du \! \left(
\frac{A_1^2}{\sqrt{(u^2-1)(1-v^2)}}\,
e^{-2\frac{\hat{F}(u)-\hat{G}(v)}{\bar{\hbar}}}\!+\frac{A_2^2}{\sqrt{(u^2-1)(1-v^2)}}\,
e^{2\frac{\hat{F}(u)-\hat{G}(v)}{\bar{\hbar}}}\right) ,
\end{gather*}
it is f\/inite if either $A_1=0$ or $A_2=0$. With our choice of sign
in $F(u)$ the fermionic ground state is:
\[
\Psi_0^{(1)}(u,v)= \frac{1}{\sqrt{u^2-v^2}}\left(
\begin{array}{c} 0\\ 0
\\e^{- \frac{ 2 (1+\delta) u + 2 (1-\delta) v}{\bar{\hbar}^2}}   \\ 0
\end{array} \right)
\]
and the norm is also given in terms of Bessel functions:
\[
N(\bar{\hbar}) = 2 \int_1^{\infty} d u \int_{-1}^1 d \bar v
{e^{-\frac{4}{\bar{\hbar}^2} (1+\delta) u} \over \sqrt{ u^2 -1} }
{e^{-\frac{4}{\bar{\hbar}^2} (1 - \delta) v} \over \sqrt{1-v^2}}=2
\pi K_0 \left( {4  \over \bar{\hbar}^2} (1+\delta) \right) I_0
\left( {4 c\over \bar{\hbar}^2} (1-\delta) \right) .
\]
It is also possible to give the fermionic ground state in ${\mathbb
R}^2$ using the ${\mathbb S}$ matrix:
\[ \Psi_0^{(1)}(x_1,x_2)={\cal S}\Psi^{(1)}_0(r_1,r_2)=\left(
\begin{array}{c} 0\vspace{1mm}\\
-\frac{(r_1+r_2)}{4 \sqrt{r_1r_2}} \sqrt{\frac{4}{r_1 r_2}
-\frac{r_1}{r_2}-\frac{r_2}{r_1}+2}
\vspace{1mm}\\ i\frac{(r_2-r_1)}{4 \sqrt{r_1r_2}} \sqrt{\frac{4}{r_1 r_2}
-\frac{r_1}{r_2}-\frac{r_2}{r_1}-2} \vspace{1mm}\\ 0
\end{array} \right)  e^{-\frac{2 ( \delta r_1
+r_2)}{\bar{\hbar}^2}}  .
\]

\begin{figure}[h]\centering
\includegraphics[height=2.5cm]{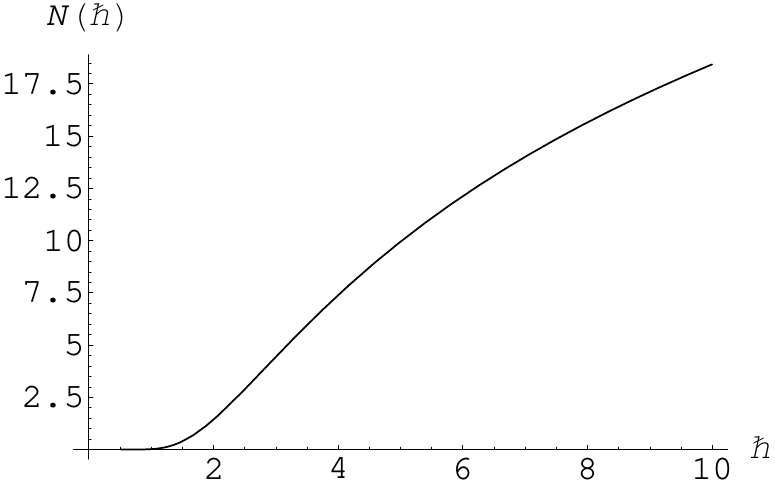}
\includegraphics[height=3.0cm]{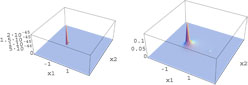}

\includegraphics[height=3.cm]{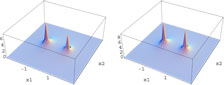}

\caption{Graphics of $N(\bar{\hbar})$ for:
 $\delta=1/2$, $\kappa=0$. $|\Psi_0^{(1)}(x_1,x_2)|^2$ for
$\delta=1/2$, $\bar{\hbar} = 0.2, \, 1, \, 4$ and $10$. Norms:
$N(0.2)= 1.3518\cdot 10^{-45}$, $N(1)=0.0178$, $N(4)=7.3881$,
$N(10)=18.4297$.}
\end{figure}

\section{The bosonic spectral problem I}\label{sec4}

The spectral problem for the scalar Hamiltonians is also separable
in elliptic coordinates. Plugging in the separation ansatz
\[
\hat{h}^{(0)\choose
(2)}\psi^{(0)\choose(2)}_E(u,v)=E\psi^{(0)\choose(2)}_E(u,v) , \qquad
\psi^{(0)\choose(2)}_E(u,v)=\eta^{(0)\choose(2)}_E(u)\xi^{(0)\choose(2)}_E(v)
\]
in the above spectral equation we f\/ind:
\begin{gather*}
\left[-\bar\hbar^2(u^2-1)\frac{d^2}{du^2}-\bar\hbar^2u\frac{d}{du}+
\left(4 \frac{(1+\delta)^2}{\bar{\hbar}^2} (u^2-1) \mp 2 (1+\delta)
u - 2 E
 u^2 \right)
\right] \eta^{(0)\choose(2)}_E(u)\\
\qquad{} =I \eta^{(0)\choose(2)}_E(u),
\\
\left[-\bar\hbar^2(1-v^2)\frac{d^2}{dv^2}+\bar\hbar^2v\frac{d}{dv}+
\left(4 \frac{(1-\delta)^2}{\bar{\hbar}^2} (1-v^2) \mp 2 (1-\delta)
v +2 E v^2 \right) \right] \xi^{(0)\choose(2)}_E(v)\\
\qquad{} = - I
\xi^{(0)\choose(2)}_E(v),
\end{gather*}
where $I$ is the eigenvalue of the symmetry operator
$\hat{I}=-\{\hat{h}^{(0)\choose (2)}+\hat{I}_2^{(0)\choose (2)}\}$.

Research on the solution of these two ODE's by power series
expansions will be published elsewhere. Here, we shall describe how
another change of variables transmutes the f\/irst ODE into Razavy
equation~\cite{Razavy1,Bondar}:
\[
-\frac{d^2\eta_{\pm}(x)}{dx^2}+ \left( \zeta_{\pm} \cosh 2x- M_\pm
\right) ^2 \eta_{\pm}(x)= \lambda_{\pm}\ \eta_{\pm}(x)  , \qquad
x={1\over 2} {\rm arccosh}\, u
\]
with parameters:
\begin{gather*}
 \zeta_{\pm}= \pm {2\over \bar{\hbar} } \sqrt{\frac{4}{\bar{\hbar}^2} (1+\delta)^2 - 2
 E_\pm}  , \qquad  \lambda_{\pm}= M_\pm^2 + {4\over \bar{\hbar}^2}
\left(I_{\pm} + 4 \frac{(1+\delta)^2}{\bar\hbar^2}\right) ,\\
M_\pm^2 ={2
(1+\delta)^2 \over 2 (1+\delta)^2 -\bar\hbar^2 E_\pm}
 \end{gather*}
Simili modo, another change of variables leads from the second ODE
to the Whittaker--Hill or Razavy trigonometric~\cite{Razavy2,Bondar}, equation
\[
 { d^2 \xi_{\pm}(y) \over d
y^2} + ( \beta_{\pm} {\rm cos} 2 y - N_\pm)^2 \ \xi_{\pm}(y) =
\mu_{\pm}\ \xi_{\pm}(y)  , \qquad y={1\over 2} \arccos v\in
\big[0,\tfrac{\pi}{2}\big]
\]
with parameters:
\begin{gather*}
\beta_{\pm} = \mp{2\over \bar{\hbar}} \sqrt{\frac{4}{\bar\hbar^2}
(1-\delta)^2 - 2 E_\pm}   , \qquad N^2_\pm = {2 (1-\delta)^2 \over 2
(1-\delta)^2 - \bar{\hbar}^2 E_\pm}  , \\ \mu_{\pm} = N^2_\pm +
{4\over \bar{\hbar}^2} \left(I_{\pm} + \frac{4}{\bar{\hbar}^2}
(1-\delta)^2\right)  .
\end{gather*}

Therefore, the spectral problem in the scalar sectors is tantamount
to the solving of two entangled sets~-- one per each pair~(E,I)~-- of
Razavy and Whittaker--Hill equations. If $M_\pm=n_1^\pm+1$,
$n_1^\pm\in {\mathbb N}^+$, the Razavy equation is QES; i.e., there
are known $n+1$ f\/inite eigenfunctions with an eigenvalue, see~\cite{FGR}:
\[E_\pm=E_{n_1^\pm} = 2
\frac{(1+\delta)^2}{\bar\hbar^2} \left(
1-\frac{1}{(n_1^\pm+1)^2}\right). \]

This means that for those values of $E_\pm$ one expects bound
eigenstates of the SUSY Hamiltonian, although the $v$-dependence
cannot be identif\/ied. If $N_\pm=n_2^\pm+1$, $n_2^\pm\in {\mathbb
N}^+$, there exist f\/inite eigenfunctions of the Whittaker--Hill
equation, with eigenvalues:
\[E_\pm=E_{n_2^\pm} = 2
\frac{(1-\delta)^2}{\bar\hbar^2} \left(
1-\frac{1}{(n_2^\pm+1)^2}\right).\]

Again, one expects  these values of $E_\pm$ to be eigenvalues of the
SUSY Hamiltonian, although their eigenfunctions are expected to be
non-normalizable (except the $n_2^+=0$ case)  because the
$u$-dependent part of the eigenfunction is non f\/inite and $u$ is a
non-compact variable. In any case, $n_1^+=n_2^+=0$ gives the bosonic
zero mode.

\subsection{The fermionic spectrum I}\label{sec4.1}

The eigenfunctions of the matrix Hamiltonian,
\begin{gather*} \hat{h}_{11}^{(1)}= -{1\over 2} \bar\hbar^2
\left( {\partial^2 \over
\partial x_1^2} + {\partial^2 \over \partial x_2^2} \right) +
{2\over\bar\hbar^2} \left[  1 +  \delta^2 + \delta \left(
{r_1\over r_2} +{r_2\over r_1} -{4  \over r_1 r_2} \right)
\right]\\
\phantom{\hat{h}_{11}^{(1)}=}{} - \left(\frac{(x_1-1)^2-x^2_2}{r_1^3} +
\delta\frac{(x_1+1)^2-x_2^2}{r_2^3} \right), \\
\hat{h}_{12}^{(1)} = \hat{h}_{21}^{(1)} = -2  \left(
\frac{x_2(x_1-1)}{r_1^3} + \delta\frac{x_2(x_1+1)}{r_2^3} \right),
\\
\hat{h}_{22}^{(1)} = -{1\over 2} \bar\hbar^2 \left( {\partial^2
\over \partial x_1^2} + {\partial^2 \over \partial x_2^2} \right) +
{2\over\bar\hbar^2} \left[  1 +  \delta^2 + \delta \left( {r_1\over
r_2} +{r_2\over r_1} -{4  \over r_1 r_2} \right) \right]
\\ \phantom{\hat{h}_{22}^{(1)} =}{}+ \left(\frac{(x_1-1)^2-x^2_2}{r_1^3} +
\delta\frac{(x_1+1)^2-x_2^2}{r_2^3} \right) ,
\end{gather*}
except the fermionic ground states, follows easily from the SUSY
algebra. The fermionic eigenfunctions in elliptic coordinates,
\begin{gather*}
\Psi^{(1)}_{E_+}(u,v)=\hat{C}_+\Psi^{(0)}_{E_+}(u,v)=\!\left(\!\!\!\begin{array}{c} 0\\\psi_{E_+}^{(1)1}(u,v)\\
\psi_{E_+}^{(1)2}(u,v)\\ 0
\end{array}\!\!\!\right)\!=-i\sqrt{\bar{\hbar}}\left(\!\!\!\begin{array}{c} 0\\e^u_1\nabla^-_u\psi_{E_+}^{(0)}(u,v)\\
e^v_2\nabla^-_v\psi_{E_+}^{(0)}(u,v)\\ 0
\end{array}\!\!\!\right) \! ,  \quad
E_+=\left\{\!\!\begin{array}{c}E_{n_1^+},\\ E_{n_2^+}, \end{array}\right.\!\!\!
\\
\Psi^{(1)}_{E_-}(u,v)=\hat{C}_-\Psi^{(2)}_{E_-}(u,v)
=\!\left(\!\!\!\begin{array}{c} 0\\\psi_{E_-}^{(1)1}(u,v)\\
\psi_{E_-}^{(1)2}(u,v)\\ 0
\end{array}\!\!\!\right)\!=-i\sqrt{\bar{\hbar}}\left(\!\!\!\begin{array}{c} 0\\-e^v_2\nabla^+_v\psi_{E_-}^{(2)}(u,v)\\
e^u_1\nabla^+_u\psi_{E_-}^{(2)}(u,v)\\ 0
\end{array}\!\!\!\right)\!  ,  \quad
E_-=\left\{\!\!\begin{array}{c}E_{n_1^-},\\ E_{n_2^-},\end{array}\right.\!\!\!
\end{gather*}
in Cartesian coordinates $\Psi_{E_\pm}^{(1)}(x_1,x_2)={\cal
S}\Psi^{(1)}_{E_\pm}(u,v)$ read:
\begin{gather*}
\Psi^{(1)}_{E_+}(x_1,x_2)=\hat{Q}_+\Psi^{(0)}_{E_+}(x_1,x_2)=\!\left(\!\!\!\begin{array}{c} 0\vspace{1mm}\\ \psi_{E_+}^{(1)1}(x_1,x_2)\vspace{1mm}\\
\psi_{E_+}^{(1)2}(x_1,x_2)\vspace{1mm}\\ 0
\end{array}\!\!\!\right)\!=-i\sqrt{\bar{\hbar}}\left(\!\!\!\begin{array}{c} 0 \vspace{1mm} \\(\bar\hbar\frac{\partial}{\partial x_1}
-\frac{\partial W}{\partial x_1})\psi_{E_+}^{(0)}(x_1,x_2)\vspace{1mm}\\
(\bar\hbar\frac{\partial}{\partial x_2}-\frac{\partial W}{\partial
x_2})\psi_{E_+}^{(0)}(x_1,x_2)\vspace{1mm}\\ 0
\end{array}\!\!\!\right)\!,
\\
\Psi^{(1)}_{E_-}(x_1,x_2)=\hat{Q}_-\Psi^{(2)}_{E_-}(x_1,x_2)=\!\left(\!\!\!\begin{array}{c} 0\vspace{1mm}\\ \psi_{E_-}^{(1)1}(x_1,x_2)\vspace{1mm}\\
\psi_{E_-}^{(1)2}(x_1,x_2)\vspace{1mm}\\ 0
\end{array}\!\!\!\right)=-i\sqrt{\bar{\hbar}}\left(\!\!\!\begin{array}{c} 0\vspace{1mm}\\ (-\bar\hbar\frac{\partial}{\partial x_2}
-\frac{\partial W}{\partial x_2})\psi_{E_-}^{(2)}(x_1,x_2)\vspace{1mm}\\
(\bar\hbar\frac{\partial}{\partial x_1}+\frac{\partial W}{\partial
x_1})\psi_{E_-}^{(2)}(x_1,x_2)\vspace{1mm}\\ 0
\end{array}\!\!\!\right)\! .
\end{gather*}

\section{Two centers of the same strength}\label{sec5}

If the two centers have the same strength, $\delta=1$, the spectral
problem in the scalar sectors becomes tantamount to the two ODE's:
\begin{gather*}
\left[-\bar\hbar^2(u^2-1)\frac{d^2}{du^2}-\bar\hbar^2u\frac{d}{du}+
\left( \frac{16}{\bar{\hbar}^2} (u^2-1) \mp 4 u - 2 E
 u^2 \right)
\right] \eta^{(0)\choose(2)}_E(u)=I \eta^{(0)\choose(2)}_E(u),
\\
\left[-\bar\hbar^2(1-v^2)\frac{d^2}{dv^2}+\bar\hbar^2v\frac{d}{dv}+
2 E v^2 \right] \xi^{(0)\choose(2)}_E(v) = - I
\xi^{(0)\choose(2)}_E(v).
\end{gather*}
Identical changes of variables as those performed in the
$0\leq\delta\leq 1$ cases now lead to the Razavy and Mathieu
equations \cite{Ince}, with parameters:
\begin{gather*}
 -\frac{d^2\eta_{\pm}(x)}{dx^2}+ \left( \zeta_{\pm} \cosh 2x- M_\pm
\right) ^2 \eta_{\pm}(x)= \lambda_{\pm}\ \eta_{\pm}(x)  ,
\\
 \zeta_{\pm}= \pm {2\over \bar{\hbar} } \sqrt{\frac{16}{\bar{\hbar}^2} - 2 E_\pm}
,
\qquad  M_\pm^2 ={8\over 8 -\bar\hbar^2 E_\pm} , \qquad
\lambda_{\pm}= M_\pm^2 + {4\over \bar{\hbar}^2} \left(I_{\pm} +
\frac{16}{\bar\hbar^2}\right) ,
\\ - { d^2 \xi_{\pm}(y) \over d y^2} + ( \alpha_\pm \cos 4 y
+ \sigma_{\pm} )\ \xi_{\pm}(y) = 0,
\qquad
\alpha_\pm =  {4 E_\pm \over \bar{\hbar}^2}  ,
\qquad   \sigma_{\pm} = {4\over \bar{\hbar}^2}  (I_{\pm} + E ) .
\end{gather*}

 We now select the three lowest energy levels from f\/inite solutions of the Razavy
equation:
\[
E_0=0 , \qquad  E_1={6\over \bar\hbar^2} , \qquad
E_2={64\over 9\bar\hbar^2}. \]

The corresponding eigenfunctions of the Razavy Hamiltonians for
$n_1^\pm=0,1,2$ are:
\begin{gather*}
 \eta_{\pm}^{01}(u) = e^{\mp {4 u \over \bar{\hbar}^2}} ,  \qquad
 \eta_{\pm}^{11} (u) = e^{\mp { 2 u \over \bar{\hbar}^2}}
\sqrt{2 (u+1)},\qquad
\eta_{\pm}^{12}(u) =- e^{\mp { 2 u \over \bar{\hbar}^2}} \sqrt{2
(u-1)},
\\  \eta_{\pm}^{21} (u) = - 2 e^{\mp {4 u \over 3
\bar{\hbar}^2}} \sqrt{ u^2 - 1} , \qquad
\eta_{\pm}^{22}(u) =\pm {3 \bar{\hbar}^2 \over 8} e^{\mp {4 u \over
3 \bar{\hbar}^2}} \left[ \pm {16 \over 3 \bar{\hbar}^2} u - 1 +
\sqrt{ 1 + { 256 \over 9
\bar{\hbar}^4}} \right],  \\
\eta_{\pm}^{23}(u) =\pm {3 \bar{\hbar}^2 \over 8} e^{\mp {4 u \over
3 \bar{\hbar}^2}} \left[ \pm {16 \over 3 \bar{\hbar}^2} u - 1 -
\sqrt{ 1 + { 256 \over 9 \bar{\hbar}^4}} \right].
\end{gather*}

The energy degeneracy is labeled by the eigenvalues of the symmetry
operator
\begin{gather*}
I^{nm}_{\pm} = \frac{\bar{\hbar}^2}{4} (
\lambda^{nm}_{\pm} - (n+1)^2) - {16\over \bar{\hbar}^2} ,
\qquad  m=1, 2, \dots, n+1,
\\
 I_{\pm}^{01} = 0  , \qquad
 I_{\pm}^{11} =  -{\bar{\hbar}^2 \over 4} - \frac{12}{\bar{\hbar}^2} \mp
 2 ,\qquad
I_{\pm}^{12} = -{\bar{\hbar}^2 \over 4} - \frac{12}{\bar\hbar^2} \pm
2,\\
I_{\pm}^{21} = - \bar{\hbar}^2 - {128 \over 9 \bar{\hbar}^2}
 , \qquad
I_{\pm}^{22} = -{\bar{\hbar}^2 \over 2} - {128\over 9 \bar{\hbar}^2
} - {1\over 6} \sqrt{ 256 + 9 \bar{\hbar}^4},\\
I_{\pm}^{23} =  -{\bar{\hbar}^2 \over 2} - {128\over 9
\bar{\hbar}^2} + {1\over 6} \sqrt{ 256 + 9 \bar{\hbar}^4}.
\end{gather*}
In this rotationally non-invariant system the symmetry operator
$\hat{I}$ replaces the orbital angular momentum in providing a basis
of common eigenfunctions with $\bar{H}$ in each degenerate in energy
sub-space of the Hilbert space in such a way that a quantum number
reminiscent of the orbital angular momentum arises.

 Next we next consider even/odd in $v$ solutions of the Mathieu equations
 \begin{gather*}
 \xi^{nm}_{\pm{\rm even}}(v) = \frac{c_1}{2} \left(  C\left[ a^{nm}_{\pm},
q_n, \arccos( v ) \right] + C\left[ a^{nm}_{\pm}, q_n, \arccos(-
v ) \right] \right)  \\
\phantom{\xi^{nm}_{\pm{\rm even}}(v) =}{} + \frac{c_2}{2} \left(  S\left[ a^{nm}_{\pm}, q_n, \arccos( v )
\right] + S\left[ a^{nm}_{\pm}, q_n, \arccos(- v ) \right] \right),
\\
 \xi^{nm}_{\pm{\rm odd}}(v) = \frac{d_1}{2} \left(  C\left[ a^{nm}_{\pm},
q_n, \arccos( v ) \right] - C\left[ a^{nm}_{\pm}, q_n, \arccos(-
v ) \right] \right)  \\
\phantom{\xi^{nm}_{\pm{\rm odd}}(v) =}{} + \frac{d_2}{2} \left(  S\left[ a^{nm}_{\pm}, q_n, \arccos( v )
\right] - S\left[ a^{nm}_{\pm}, q_n, \arccos(- v ) \right] \right).
\end{gather*}
The reason is that the invariance of the problem under
$r_1\leftrightarrow r_2\equiv v\leftrightarrow -v$ -- the exchange
between the two centers~-- forces even or odd eigenfunctions in $v$.
The parameters of the Mathieu equations determined by the spectral
problem are:
\[
a^{nm}_{\pm}  = - \frac{\sigma^{nm}_{\pm}}{4} = -\frac{E_n +
I^{nm}_{\pm}}{\bar{\hbar}^2}  ,  \qquad q_n =
\frac{\alpha_n}{8}= \frac{E_n} {2\bar{\hbar}^2}  .
\]
To f\/it in with the parameters of the Razavy Hamiltonians set by
$n_1^\pm=0,1,2$ we must choose: \begin{gather*}
 q_0 = 0  , \qquad a^{01}_{\pm} = 0,  \\
 q_1 = \frac{3}{\bar{\hbar}^4} , \qquad
 a^{11}_{\pm} = {6 \over \bar{\hbar}^4} \pm\frac{
2}{\bar{\hbar}^2} + \frac{1}{4}, \qquad a^{12}_{\pm} = {6 \over
\bar{\hbar}^4} \mp \frac{ 2}{\bar{\hbar}^2} + \frac{1}{4}, \\
 q_2 = \frac{32}{9 \bar{\hbar}^4} , \qquad
a^{21}_{\pm} = 1 + {64 \over 9 \bar{\hbar}^4} ,  \qquad
 a^{22}_{\pm} =\frac{1}{2} + {64 \over 9\bar{\hbar}^4} +
{1\over 6 \bar{\hbar}^2 } \sqrt{256 + 9 \bar{\hbar}^4}, \\
a^{23}_{\pm} = {1 \over 2} + {64 \over 9\bar{\hbar}^4} - {1\over
6\bar{\hbar}^2} \sqrt{ 256 + 9 \bar{\hbar}^4} .
\end{gather*}
Therefore,
\[ \psi^{(0)}_{nm\,\,{\rm even/odd}}(u,v) = \eta^{nm}_+
(u) \xi^{nm}_{+{\rm even/odd}}(v) , \qquad n\geq 0, \qquad
m=1,2, \dots ,m+1  ,
\]
is a set of bound states of non-zero energy of the scalar
Hamiltonian of two SUSY Coulombian centers of the same strength. The
paired fermionic eigenstates are obtained through the action of
the appropriate supercharge. Since every non-zero-energy state come
in bosonic-fermionic pairs, the criterion for spontaneous
supersymmetry breaking is the existence of a~Fermi--Bose pair ground
state of positive energy connected one with each other by one of the
supercharges. Our SUSY Hamiltonian has both bosonic and fermionic
zero modes as single ground states; consequently, supersymmetry is
not spontaneously broken in this system.

In the next f\/igures we show several graphics of bosonic SUSY
eigenfunctions for the choice: $c_2=d_2=0$, $c_1=d_1=1$.
\begin{figure}[h]\centering
\includegraphics[height=3cm]{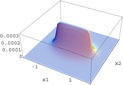} \
\includegraphics[height=3cm]{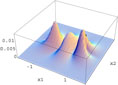}

\includegraphics[height=3cm]{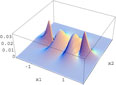} \
\includegraphics[height=3cm]{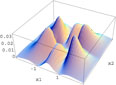} \
\includegraphics[height=3cm]{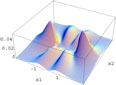}

\caption{Graphics with ($\bar{\hbar}=1$):
$|\psi^{01}_{+{\rm even}}(x_1,x_2)|^2$; $|\psi^{11}_{+{\rm
even}}(x_1,x_2)|^2$  and $|\psi^{11}_{+{\rm odd}}(x_1,x_2)|^2$;
$|\psi^{21}_{+{\rm even}}(x_1,x_2)|^2$ and $|\psi^{21}_{+{\rm
odd}}(x_1,x_2)|^2$.}
\end{figure}

\subsection{Bosonic and fermionic ground states for two centers of the same strength}\label{sec5.1}

For comparison, we also plot the bosonic and fermionic ground states
for several values of $\bar\hbar$.
\begin{figure}[h]\centering
\includegraphics[height=2.75cm]{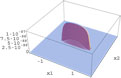}
\includegraphics[height=2.75cm]{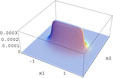}
\includegraphics[height=2.75cm]{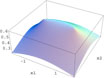}
\includegraphics[height=2.75cm]{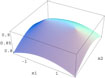}

\caption{Graphics of the probability
density $|\Psi_0^{(0)}(x_1,x_2)|^2$ for $\delta=1$, and the values
of $\bar{\hbar} = 0.2, \, 1, \, 4$ and $10$.}
\end{figure}

\begin{figure}[h]
\includegraphics[height=2.7cm]{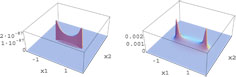}
\includegraphics[height=2.7cm]{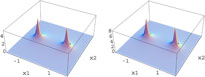}

\caption{Graphics of
$|\Psi_0^{(1)}(x_1,x_2)|^2$ for $\delta=1$, and $\bar{\hbar} = 0.2,
\, 1, \, 4$ and $10$. The norms are: $N(0.2)= 7.7012\cdot 10^{-88}$,
$N(1)=0.0009$, $N(4)=5.8083$, $N(10)=16.6347$.}
\end{figure}

\section{The superpotential II for the two-center problem}\label{sec6}

There is another possibility to f\/ind the potential energy of two
Coulombian centers in a 2D ${\cal N}=2$ SUSY Hamiltonian. The
superpotential must satisfy the Hamilton--Jacobi equation rather than
the Poisson equation (\ref{eq:sups}):
\begin{equation} \frac{ 1}{r_1}+\frac{\delta}{r_2}=\frac{1}{2}
\left( \frac{\partial W}{\partial x_1}\right)^2+\frac{1}{2} \left(
\frac{\partial W}{\partial x_2}\right)^2.\label{eqHJ}
\end{equation}
Note that in this case the two centers must be repulsive to
guarantee a real $W$. This point of view, which follows the path
shown in \cite{Manton} and \cite{Heumann} for the Coulomb problem,
amounts to the quantization of a classical supersymmetric system,
only semi-positive def\/inite for repulsive potentials.

Again using elliptic coordinates, the Hamilton--Jacobi equation
separates
\[ \kappa =-(u^2-1) \left({dF\over d u}\right)^2 + 2 (1+\delta) u
, \qquad \kappa = (1-v^2) \left({dG\over d v}\right)^2 +2
(\delta-1) v \label{kappa}
\]
by plugging in (\ref{eqHJ}) the ansatz:
\[
W(u,v;\kappa)=F_a(u;\kappa) + G_b(v;\kappa)   , \qquad  a,b=0,1
.
\]
The quadratures
\[
F_a(u;\kappa)= (-1)^a\, \int_{1}^u \frac{\sqrt{2 (1+\delta)
u-\kappa}}{\sqrt{u^2-1}}\, du  , \qquad G_b(v;\kappa)= (-1)^b\,
\int_{-1}^v \frac{\sqrt{2 (1-\delta) v+\kappa}}{\sqrt{1-v^2}}\, dv
\]
show that the separation constant $\kappa$ is constrained in order
to f\/ind real $F_a(u)$ and $G_b(v)$: $2(1-\delta) \leq \kappa \leq 2
(1+\delta)$. Note that there are two dif\/ferent possibilities: $a=b$
and $a\neq b$. A~dif\/ferent global sign only exchanges the $(0)$ with
the $(2)$ and the $(1)1$ with the $(1)2$ sectors. The superpotential
II is thus given in terms of incomplete and complete elliptic
integrals of the f\/irst and second type, see \cite{Abramowitz,Byrd}:
\begin{gather*} F_a(u;\kappa) = (-1)^a 2 i
\sqrt{\kappa+2(1{+}\delta) }  \left( E\left[ \sin^{-1}
\sqrt{\frac{\kappa-2(1{+}\delta)u}{\kappa-2(1{+}\delta)
}},\frac{\kappa-2(1{+}\delta)}{\kappa+2(1{+}\delta) }\right]\right.\\ \left.
{} -E\left[
{\pi\over 2},\frac{\kappa-2(1{+}\delta)}{\kappa+2(1{+}\delta)
}\right]-F\left[ \sin^{-1}
\sqrt{\frac{\kappa-2(1{+}\delta)u}{\kappa-2(1{+}\delta)}},\frac{\kappa-2(1{+}\delta)}{\kappa+2(1{+}\delta)
}\right]+F\left[{\pi\over
2},\frac{\kappa-2(1{+}\delta)}{\kappa+2(1{+}\delta)}\right]\right),
\\ G_b(v;\kappa)  =  (-1)^b 2 i
\sqrt{\kappa-2(1{-}\delta) }\\
{}\times\left(- E\left[ \sin^{-1}
\sqrt{\frac{\kappa+2(1{-}\delta)v}{\kappa+2(1{-}\delta)
}},\frac{\kappa+2(1{-}\delta)}{\kappa-2(1{-}\delta)}\right]+E\left[ {\rm
sin}^{-1}\sqrt{\frac{\kappa-2(1{-}\delta)}{\kappa+2(1{-}\delta)}},\frac{\kappa+2(1{-}\delta)}{\kappa-2(1{-}\delta)
}\right]\right. \\  {} +\left. F\left[ \sin^{-1}
\sqrt{\frac{\kappa+2(1{-}\delta)v}{\kappa+2(1{-}\delta)}},\frac{\kappa+2(1{-}\delta)}{\kappa-2(1{-}\delta)
}\right] - F\left[{\rm
sin}^{-1}\sqrt{\frac{\kappa-2(1{-}\delta)}{\kappa+2(1{-}\delta)}},\frac{\kappa+2(1{-}\delta)}{\kappa-2(1{-}\delta)
}\right]\right).
\end{gather*}
The scalar Hamiltonians, both in elliptic and Cartesian coordinates,
read:
\begin{gather*} \hat{h}^{(0)\choose(2)}={1\over
2(u^2-v^2)}\Bigg\{-\bar\hbar^2\left((u^2-1)\frac{d^2}{du^2}+u\frac{d}{du}+
(1-v^2)\frac{d^2}{dv^2}- v\frac{d}{dv}\right) \\
\phantom{\hat{h}^{(0)\choose(2)}=}{}+
2(1+\delta)u+2(1-\delta)v  \\
\phantom{\hat{h}^{(0)\choose(2)}=}{}
\pm\bar\hbar\left[(-1)^a(1+\delta)\sqrt{\frac{u^2-1}{2(1+\delta)u-\kappa}}+
(-1)^b(1-\delta)\sqrt{\frac{1-v^2}{2(1-\delta)v+\kappa}}
\right] \Bigg\},
\\
\hat{h}^{(0)\choose(2)}=-{\bar\hbar^2\over
2}\nabla^2 +\frac{1}{r_1}+\frac{\delta}{r_2}\\
\phantom{\hat{h}^{(0)\choose(2)}=}{} \pm\frac{\bar{\hbar}}{
4 r_1 r_2} \left\{  \frac{(-1)^a (1+\delta) \sqrt{(r_1+r_2)^2\!-4
}}{\sqrt{(1+\delta) (r_1+r_2)-\kappa}} + \frac{(-1)^b (1-\delta)
\sqrt{4-(r_1-r_2)^2}}{\sqrt{ \kappa-(1-\delta) (r_1-r_2)}} \right\} .
\end{gather*}

\subsection{Type IIa and Type IIb two-center SUSY quantum mechanics}\label{sec6.1}

 We now present the graphics of the scalar potential for $a=b$, a
 system that we shall call Type~IIa ${\cal N}=2$ SUSY two Coulombian
 centers. By the same token, the system arising from the
 superpotential I will be called Type I ${\cal N}=2$ SUSY two Coulombian
 centers.

\begin{figure}[h]\centering
\includegraphics[height=4cm]{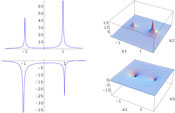}\
\includegraphics[height=4cm]{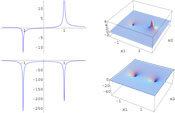}

\caption{3D graphics of the quantum
potential $\hat{V}^{(0)}$ for ${\bf a}={\bf b}={\bf 1}$ (or $\hat{V}^{(2)}$ for
${\bf a}={\bf b}={\bf 0}$). We choose $\delta=1/2$, $\kappa=3$. Cases: (a)
$\bar{\hbar}=0.2$, $\bar{\hbar}= 2$. Observe that when $\bar\hbar$
is increased this provides a reduction of the strength of the
repulsive centers and the left center becomes attractive whereas the
right center is still repulsive. (b) $\bar{\hbar}=4$ and
$\bar{\hbar}=10$, both centers are attractive. With increasing
$\bar{\hbar}$ the centers become more and more attractive.}
\end{figure}

If $a\neq b$, we shall call the system Type IIb ${\cal N}=2$ SUSY
two Coulombian centers. The scalar potential is drawn in the
graphics below for several values of $\bar\hbar$.

\begin{figure}[h]\centering
\includegraphics[height=4.8cm]{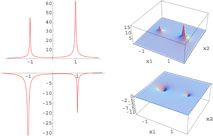} \
\includegraphics[height=4.8cm]{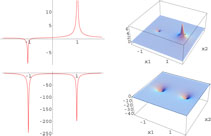}

\caption{3D graphics of the quantum
potential $\hat{V}^{(0)}$ for ${\bf a}={\bf 1}$, ${\bf b}={\bf 0}$ (or
$\hat{V}^{(2)}$ for ${\bf a}={\bf 0}$, ${\bf b}={\bf 1}$). We choose $\delta=1/2$,
$\kappa=3$. Cases: (a) $\bar{\hbar}=0.2$,  $\bar{\hbar}= 2$. Observe
that once again increasing $\bar{\hbar}$ provides a~reduction of the
strength of the repulsive centers and the left center becomes
attractive whereas the right center is still repulsive. (b)
$\bar{\hbar}=4$ and $\bar{\hbar}=10$, both centers are attractive.}
\end{figure}

The dif\/ferences between the Type IIa and Type IIb potentials,
increasing with $\bar\hbar$, are shown in the next f\/igure.

\begin{figure}[h]\centering
\includegraphics[height=4.2cm]{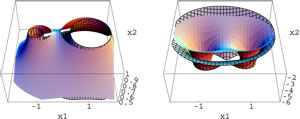}

\caption{3D graphics of the quantum
potential $\hat{V}^{(0)}$ for ${\bf a}={\bf 1}={\bf b}$ and ${\bf a}={\bf 1}$, ${\bf b}={\bf 0}$
in the cases $\bar{\hbar}=1$,  $\bar{\hbar}= 10$. In this range we
note the dif\/ferences between Type IIa and IIb quantum potentials.}
\end{figure}

Below we shall discuss the coincidences and dif\/ferences of the three
distinct points of view.

\subsection{The spectral problem}\label{sec6.2}

Both for the Type IIa and Type IIb systems the spectral problem in
the scalar sectors is separable in elliptic coordinates:
\begin{gather*} \hat{h}^{(0)\choose
(2)}\psi^{(0)\choose(2)}_E(u,v)=E\psi^{(0)\choose(2)}_E(u,v) ,  \qquad
\psi^{(0)\choose(2)}_E(u,v)=\eta^{(0)\choose(2)}_E(u)\zeta^{(0)\choose(2)}_E(v),\\
\left[-\bar\hbar^2(u^2-1)\frac{d^2}{du^2}-\bar\hbar^2u\frac{d}{du}+
2(1+\delta)u-2Eu^2\pm\bar\hbar(-1)^a(1+\delta)\sqrt{\frac{u^2-1}{2(1+\delta)u-\kappa}}\,
\right]\!\eta^{(0)\choose(2)}_E(u)\!\!\\
\qquad{}  =I\eta^{(0)\choose(2)}_E(u),
\\
\left[-\bar\hbar^2(1-v^2)\frac{d^2}{dv^2}+\bar\hbar^2v\frac{d}{dv}+
2(1-\delta)v+2Ev^2\pm\bar\hbar(-1)^b(1-\delta)\sqrt{\frac{1-v^2}{2(1-\delta)v-\kappa}} \,\right]\!
\zeta^{(0)\choose(2)}_E(v)\!\!\\
\qquad{} =-I\zeta^{(0)\choose(2)}_E(v) .
\end{gather*}
The separated ODE's are, however, much more dif\/f\/icult (non-linear)
than in the Type I system and there is no hope of f\/inding explicit
eigenvalues and eigenfunctions.

\subsection{Bosonic ground states}\label{sec6.3}

We shall therefore concentrate on searching for the ground states.
First, the bosonic zero modes:
\[
\hat{C}_+\Psi_0^{(0)}(u,v)=0, \qquad
\hat{C}_-\Psi_0^{(2)}(u,v)=0 .
\]
The separation ansatz
\[
\psi^{(0)\choose
(2)}_0(u,v)=\eta^{(0)\choose(2)}_0(u)\zeta^{(0)\choose(2)}_0(v)
\]
makes these equations equivalent to:
\begin{gather*}
e^u_1\nabla_u^- \eta_0^{(0)}(u)=0 , \qquad e^u_1\nabla_u^+
\eta_0^{(2)}(u)=0, \qquad e^v_2\nabla_v^- \zeta_0^{(0)}(v)=0 ,
\qquad -e^v_2\nabla_v^+ \zeta_0^{(2)}(v)=0 ,
\end{gather*}
or,
\begin{gather*}
\left(\bar\hbar\frac{d}{du}\mp(-1)^a\sqrt{\frac{2(1+\delta)u-\kappa}{u^2-1}}\right)\eta^{(0)\choose(2)}_0(u)=0
, \\
\left(\bar\hbar\frac{d}{dv}\mp(-1)^b\sqrt{\frac{2(1-\delta)v+\kappa}{1-v^2}}\right)\zeta^{(0)\choose(2)}_0(v)=0
.
\end{gather*}
The solutions, i.e.\ the bosonic zero modes, are:
\[
\eta_0^{(0)\choose(2)}(v)=\exp\left[\pm
\frac{F_a(u;\kappa)}{\bar\hbar}\right], \qquad
\zeta_0^{(0)\choose(2)}(v)=\exp \left[\pm
\frac{G_b(v;\kappa)}{\bar\hbar}\right].
\]
It is not possible to calculate analytically the norm in these
cases, but we of\/fer a numerical integration of the $F=0$ bosonic
ground states:
\[ N(\bar{\hbar};\kappa,a,b)=2\int_{-1}^1\, dv\int_1^{\infty}\,
du\, \frac{u^2-v^2}{\sqrt{u^2-1}\, \sqrt{1-v^2}} \exp \left[\frac{2 \, F_a(u;\kappa)}{\bar{\hbar}}\right]\exp \left[\frac{2 \, G_b(v;\kappa)}{\bar{\hbar}}\right] \] in the
next f\/igures. We observe that there is one normalizable bosonic
ground state of zero energy of Type IIa and one of Type IIb once the
value of $a$ is set to be one. Remarkably, the Type IIb zero mode
disappears (the norm becomes inf\/inity) at the classical limit.

\begin{figure}[h]\centering
\begin{minipage}[b]{75mm}\centering
\includegraphics[height=4cm]{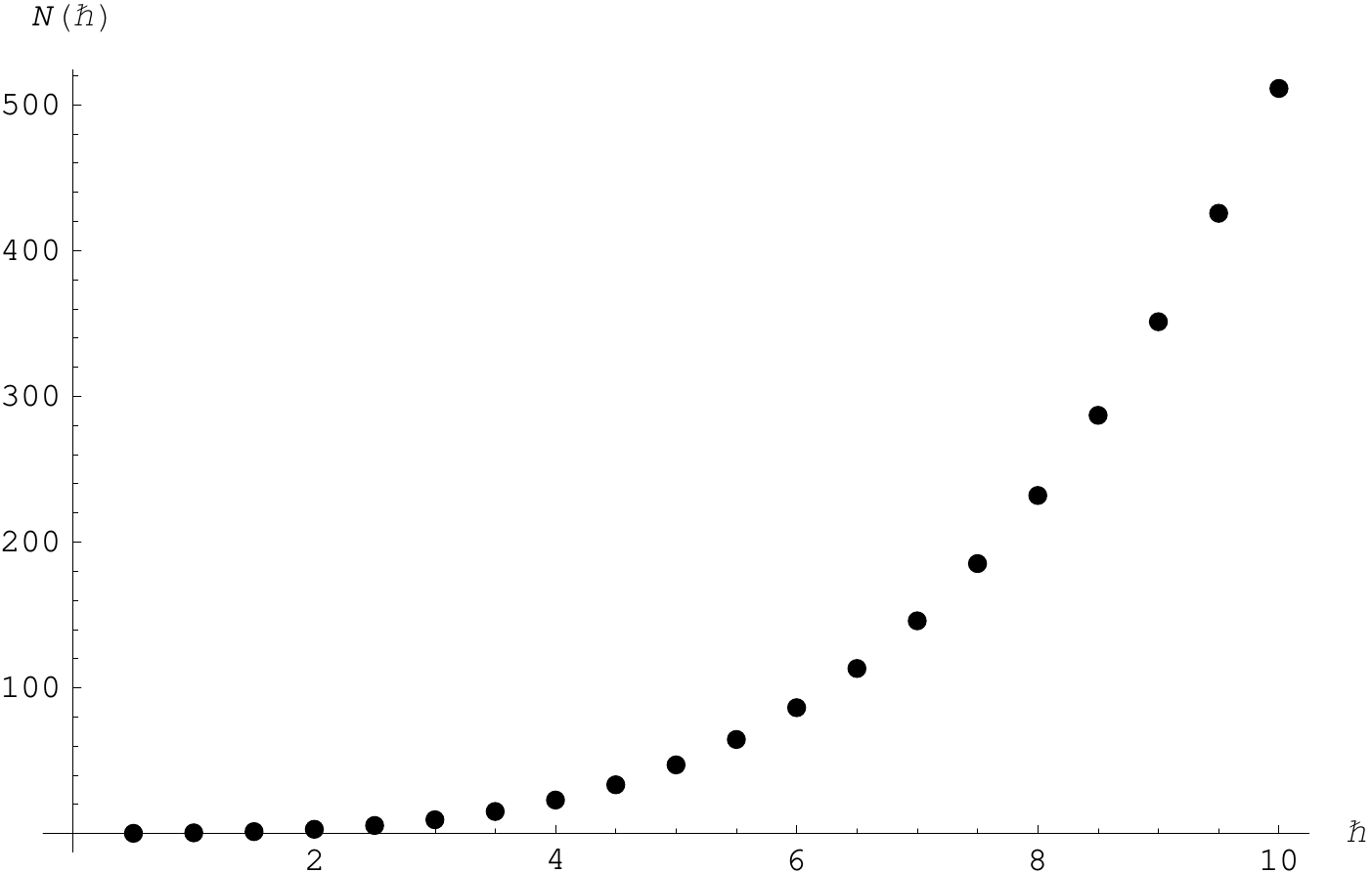}
\caption{Numerical plot of
$N(\bar{\hbar};3,1,1)$ as function of $\bar{\hbar}$ for
$\delta=1/2$.}
\end{minipage}\qquad
\begin{minipage}[b]{75mm}\centering
\includegraphics[height=4cm]{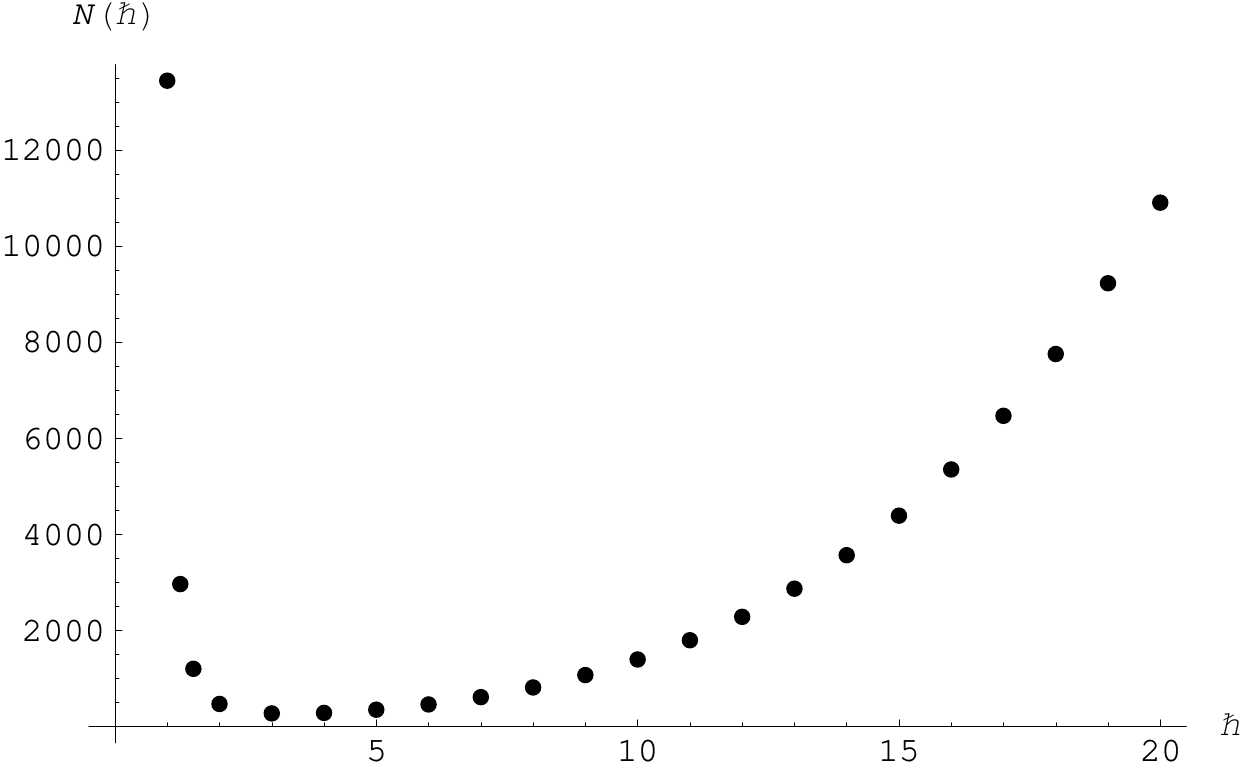}
\caption{Numerical
plot of $N(\bar{\hbar};3,1,0)$ as function of $\bar{\hbar}$ for
$\delta=1/2$.}
\end{minipage}
\end{figure}

Some 3D plots of the Type IIa and Type IIa bosonic zero modes for
several values of $\bar\hbar$ are shown in the next f\/igures:

\begin{figure}[h]\centering
\includegraphics[height=2.7cm]{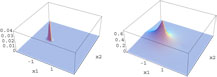} \ \includegraphics[height=2.7cm]{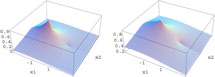}
\caption{3D graphics of the ground state
probability density $|\Psi^{(0)}(x_1,x_2)|^2$, for $\delta=1/2$,
$\kappa=3$, and ${\bf a}={\bf b}={\bf 1}$ (or $|\Psi^{(2)}(x_1,x_2)|^2$ for
${\bf a}={\bf b}={\bf 0}$). Cases: $\bar{\hbar}=0.2$, $\bar{\hbar}= 2$,
$\bar{\hbar}=4$ and $\bar{\hbar}=10$. The norms are
$N(0.2)=0.004914$, $N(2)=2.83912$, $N(4)=22.8914$ and
$N(10)=511.092$.}
\end{figure}

\begin{figure}[h]\centering
\includegraphics[height=2.7cm]{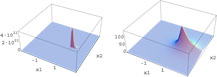} \
\includegraphics[height=2.7cm]{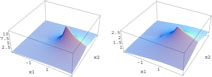}

\caption{3D graphics of the ground state
probability density  $|\Psi^{(0)}(x_1,x_2)|^2$, for $\delta=1/2$,
$\kappa=3$, and ${\bf a}={\bf 1}$, ${\bf b}={\bf 0}$ (or $|\Psi^{(2)}(x_1,x_2)|^2$
for ${\bf a}={\bf 0}$, ${\bf b}={\bf 1}$). Cases: $\bar{\hbar}=0.2$, $\bar{\hbar}=
2$, $\bar{\hbar}=4$ and $\bar{\hbar}=10$. And the norms are $N(0.2)
= 9.61622 \cdot 10^{20} $, $N(2)=473.903$, $N(4)=287.687$ and
$N(10)=1399.71$.}
\end{figure}

\subsection{Fermionic ground states}\label{sec6.4}

Second, the fermionic ground states of zero energy:
\[
\hat{C}_+\Psi_0^{(1)}(u,v)=0 , \qquad
\hat{C}_-\Psi_0^{(1)}(u,v)=0 .
\]
Unlike the bosonic case where the logic is {\it or} instead
of {\it and}, note that both equations must be satisf\/ied by
the fermionic zero modes. The separation ansatz
\[
\psi_0^{(0)1}(u,v)=\eta_0^{(0)1}(u)\zeta_0^{(0)1}(v) , \qquad
\psi_0^{(0)2}(u,v)=\eta_0^{(0)2}(u)\zeta_0^{(0)2}(v)
\]
makes these equations tantamount to
\begin{gather*}
 e^u_1\left(\nabla_u^++\frac{\bar\hbar u}{u^2-v^2}\right)
\psi_0^{(1)1}(u,v)+e^v_2\left(\nabla_v^+-\frac{\bar\hbar v}{u^2-v^2}\right) \psi_0^{(1)2}(u,v)=0, \\
 -e^v_2\left(\nabla_v^--\frac{\bar\hbar v}{u^2-v^2}\right)
\psi_0^{(1)1}(u,v)+e^u_1\left(\nabla_u^-+\frac{\bar\hbar u}{u^2-v^2}\right)
\psi_0^{(1)2}(u)=0 ,
\end{gather*}
or,
\begin{gather*}
 \bar\hbar\frac{d\eta_0^{(1)1}}{du}+\left(\frac{d
F_a}{du}+\frac{\bar\hbar u}{u^2-v^2}\right)\eta_0^{(1)1}(u)=0
, \qquad \bar\hbar\frac{d\eta_0^{(1)2}}{du}-\left(\frac{d
F_a}{du}-\frac{\bar\hbar u}{u^2-v^2}\right)\eta_0^{(1)2}(u)=0,
\\ \bar\hbar\frac{d\zeta_0^{(1)1}}{dv}-\left(\frac{d
G_a}{dv}+\frac{\bar\hbar v}{u^2-v^2}\right)\zeta_0^{(1)1}(v)=0
, \qquad \bar\hbar\frac{d\zeta_0^{(1)2}}{dv}+\left(\frac{d
G_a}{dv}-\frac{\bar\hbar v}{u^2-v^2}\right)\zeta_0^{(1)2}(v)=0  .
\end{gather*}

The fermionic zero modes have the form of the linear combination:
\begin{gather*}
\Psi_0^{(1)}(u,v)=\frac{1}{\sqrt{u^2-v^2}}\left\{A_1\left(
\begin{array}{c} 0\\
\exp [-\frac{(F_a(u;\kappa)-G_b(v;\kappa)}{\bar{\hbar}}]\\ 0 \\
0
\end{array} \right) +  A_2\left(
\begin{array}{c} 0\\0\\
\exp [\frac{F_a(u;\kappa)-G_b(v;\kappa)}{\bar{\hbar}}]\\
0
\end{array} \right)\right\}.
\end{gather*}
Because the norm is
\[
N(\bar{\hbar};\kappa,a,b)=2\int_{1}^{\infty}\int_{-1}^1 \frac{du
dv}{\sqrt{(u^2-1)(1-v^2)}}   \left( A_1^2
e^{-2\frac{F_a(u;\kappa)-G_b(v;\kappa)}{\bar{\hbar}}}+A_2^2
e^{2\frac{F_a(u;\kappa)-G_b(v;\kappa)}{\bar{\hbar}}}\right)  .
\]
only $A_1=0$ or $A_2=0$ are normalizable, depending on the choice of
$a$ and $b$.

Again, these integrals cannot be computed analytically but the
outcome of numerical calculations is shown in the next f\/igures for
several values of $\bar\hbar$.
\begin{figure}[h]\centering
\includegraphics[height=3.5cm]{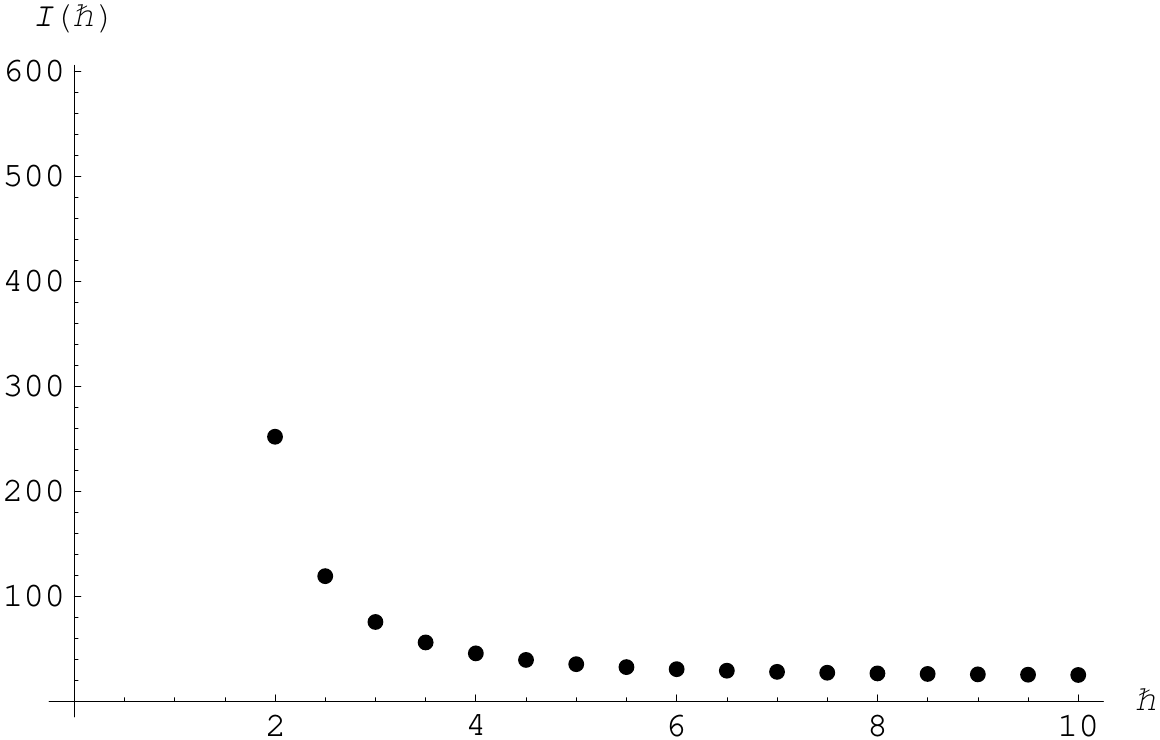} \qquad\quad \includegraphics[height=3.5cm]{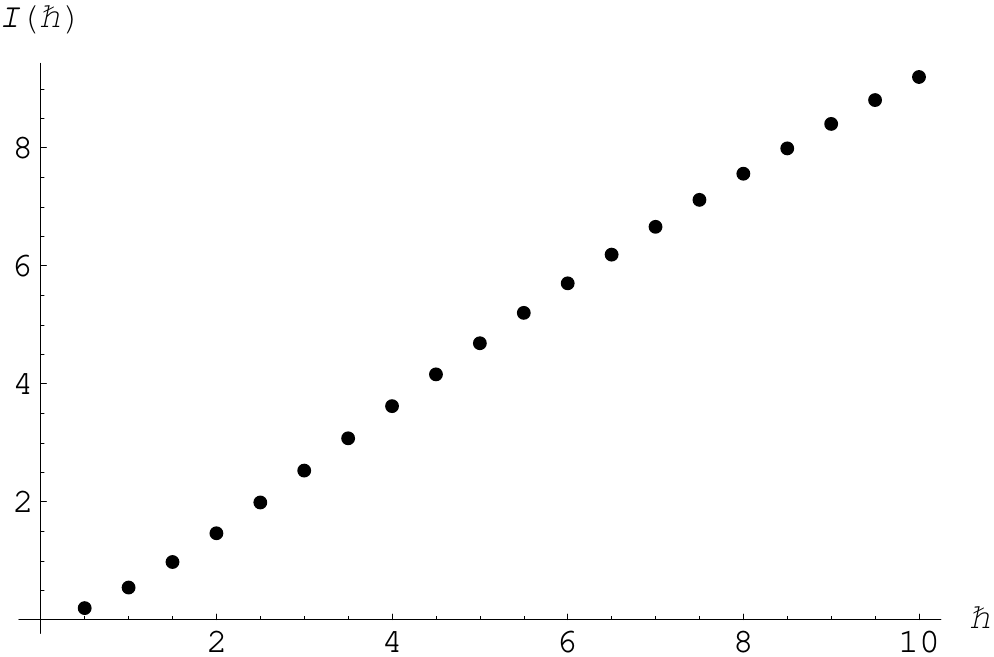}
\caption{Numerical plots of
$N(\bar{\hbar};3,1,1)$ and $N(\bar{\hbar;3,1,0})$ for $A_1=1$,
$A_2=0$ and $\delta=1/2$ ($N(\bar{\hbar};3,0,0)$ and
$N(\bar{\hbar;3,0,1})$ for $A_1=0$, $A_2=1$).}
\end{figure}

In this case we see that the fermionic zero mode of Type IIa does
not have a classical limit whereas the Type IIb fermionic ground
state behaves smoothly near $\bar\hbar=0$.

3D plots of fermionic zero modes, of both Type IIa and IIb, are
shown in the last f\/igures.
\begin{figure}[h]\centering
\includegraphics[height=2.65cm]{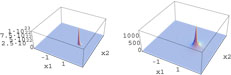}\
\includegraphics[height=2.65cm]{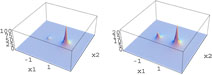}

\caption{Graphics of
$|\Psi_0^{(1)(1)}(x_1,x_2)|^2$ for $A_1=1$, $A_2=0$, $\delta=1/2$,
$\kappa=3$, and the sign combination: ${\bf a}={\bf b}={\bf 1}$. Cases:
$\bar{\hbar}=0.2, 2, 4$ and $10$. These graphics also represent
$|\Psi_0^{(1)(2)}(x_1,x_2)|$ with $A_1=0$, $A_2=1$, $\delta=1/2$,
$\kappa=3$, and ${\bf a=b=0}$. The norms are $N(0.2)= 1.03792 \cdot
10^{22} $, $N(2)=251.908$, $N(4)=45.7495$ and $N(10)=25.207$.}

\includegraphics[height=2.7cm]{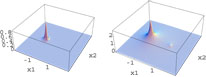}\
\includegraphics[height=2.7cm]{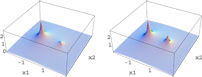}

\caption{Graphics of the function
$|\Psi^{(1)(1)}(x_1,x_2)|^2$ for $A_1=1$, $A_2=0$, $\delta=1/2$,
$\kappa=3$, and the sign combination: ${\bf a}={\bf 1}$, ${\bf b}={\bf 0}$. Cases:
$\bar{\hbar}=0.2, 2, 4$ and $10$. These graphics also represent
$|\Psi_0^{(1)(2)}(x_1,x_2)|^2$ for $A_1=0$, $A_2=1$ and ${\bf a}={\bf 0}$,
${\bf b}={\bf 1}$. The norms are $N(0.2)=0.05046$, $N(2)=1.46657$,
$N(4)=3.62192$ and $N(10)=9.20207$.}
\end{figure}

\section{Summary}\label{sec7}

In this paper we have built and studied two types of supersymmetric
quantum mechanical systems starting from two Coulombian centers of
force. Our theoretical analysis could be of interest in molecular
physics seeking supersymmetric spectra closely related to the
spectra of homonuclear or heteronuclear diatomic molecular ions,
e.g., the hydrogen molecular ion or the same system with a proton
replaced by a deuteron. In $H_2^+$ for instance, the f\/irst case, the
value of the non-dimensional quantization parameter is checked to be
$\bar\hbar=0.7$:
\[
\hbar = 1.05 \cdot 10^{-34} \ {\rm kg} \cdot {\rm m}^2 \cdot {\rm
s}^{-1} , \qquad  \sqrt{m d \alpha} = 1.493 \cdot
10^{-34} \ {\rm kg }\cdot{\rm m}^2 \cdot {\rm s}^{-1} ,
\]
using the international system of units (SI).

The f\/irst type is def\/ined from a superpotential that solves the
Poisson equation with the potential of the two centers as the
source. In this case, the spectral problem is shown to be equivalent
to entangled families of Razavy and Whittaker--Hill equations. Using
the property of the quasi-exact solvability of these systems, many
eigenvalues of the SUSY system corresponding to bound states have
been identif\/ied when the strengths of the two centers are dif\/ferent.
If the strengths are equal things become easier and some bound
states are also found. In summary, for our simplest choice of Type~I
superpotential the main features of the spectrum are the following:
\begin{itemize}\itemsep=0pt

\item There is an inf\/inite set of discrete energy eigenvalues in the
$F=0$ Bose sub-space of the Hilbert space:
\[
E_n=2\frac{m\alpha^2}{\hbar^2}
(1+\delta)^2\left(1-\frac{1}{(n+1)^2}\right) .
\]
The ionization energy is: $E_\infty =2\frac{m\alpha^2}{\hbar^2}
(1+\delta)^2$, the threshold of the continuous spectrum.

\item There is a sub-space of dimension $n+1$ of degenerate
eigenfunctions with energy $E_n$. The eigenvalues of the symmetry
operator $\hat{I}$
\[
I_{nm}=\frac{\hbar^2}{4}\left((\lambda_{nm}-(n+1)^2\right) ,
\qquad m=1,2, \dots , n+1 ,
\]
where $\lambda_{nm}$ are the $n+1$ roots of the polynomial that
solves the Razavy equation, label a~basis of eigefunctions in each
energy sub-space.

\item In the case $\delta=1$ the eigenfunctions can be
explicitly found. Moreover, the system enjoys a discrete symmetry
under center exchange: $v\leftrightarrow -v$ $(r_1\leftrightarrow
r_2)$. The energy eigenfunctions come in even/odd pairs of functions
of $v$ with respect to this ref\/lection. This fact suggests that the
purely bosonic two f\/ixed centers Hamiltonian
\[
H=-{1\over 2m}\nabla^2+{\alpha\over r_1}+{\alpha\over r_2}+C
\]
enjoys a hidden supersymmetry (if the constant $C$ is greater than
$8\frac{m\alpha^2}{\hbar^2}$) of the kind recently unveiled in
\cite{Ply} for classy one-dimensional systems. This hidden
supersymmetry is spontaneously broken because the even/odd ground
states have positive energy:
$E_\pm=C-8\frac{m\alpha^2}{\hbar^2}>0$.

\item There are eigenfunctions in the $F=1$ Fermi sub-space for
the same eigenvalues $E_n$, $n>0$. Analytically, the Fermi
eigenfunctions are obtained from the Bose eigenfunctions through the
action of $\hat{Q}_+$.

\end{itemize}

The second type starts from superpotentials solving the
Hamilton--Jacobi equation. There are two non-equivalent sign
combinations giving two sub-classes. Both Type IIa and Type IIb
superpotentials are def\/ined in terms of incomplete and complete
elliptic integrals of the f\/irst and second kind. The superpotential
in this approach is no more than the Hamilton's characteristic
function for zero energy and f\/lipped potential. The separability of
the HJ equation in elliptic coordinates means that we can f\/ind a
``complete'' solution of this equation. The spectral problem is,
however, hopeless for this Type.

All the zero-energy ground states, bosonic and fermionic, Type I and
Type IIa/IIb, dif\/ferent strengths and equal strengths, have been
obtained. The cross section ($x_2=0$) of the probability density of
some of the ground states are shown, for the sake of comparison, in
these f\/inal Tables. It is remarkable that, despite of being
analytically very dif\/ferent, Type I and Type II zero modes show
similar patterns.

\begin{table}[htdp]\centering \caption{$\bar{\hbar} = 1$, $\delta=1/2$.}
\vspace{1mm}

\begin{tabular}{|@{}c@{}|@{}c@{\,\,}c@{\,\,}c@{}|}  \hline  \tsep{2ex}
$|\Psi_{0}^{(0/1)}(x_1,x_2)|^2$ &  Type I  &  $\begin{array}{@{}c@{}} {\rm Type \ IIa}\\ \kappa=3\end{array}$
  &  $\begin{array}{@{}c@{}} {\rm Type \ IIb} \\ \kappa=3 \end{array}$ \bsep{1ex} \\
 \hline \tsep{2ex} $\begin{array}{@{}c@{}} {\rm Bosonic}\\ {\rm zero \ mode}\end{array}$   &
\parbox{4cm}{\includegraphics[width=4cm]{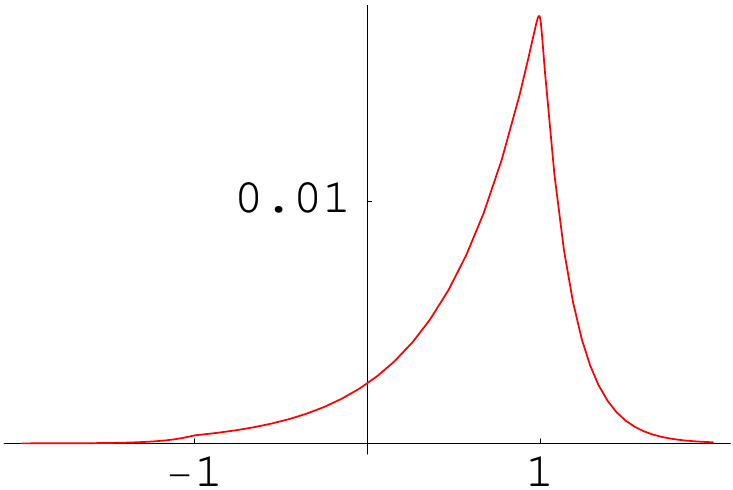}} &
\parbox{4cm}{\includegraphics[width=4cm]{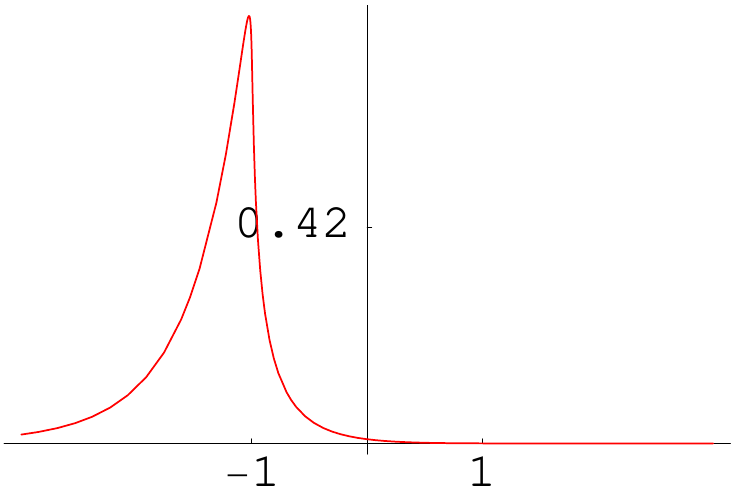}} &
\parbox{4cm}{\includegraphics[width=4cm]{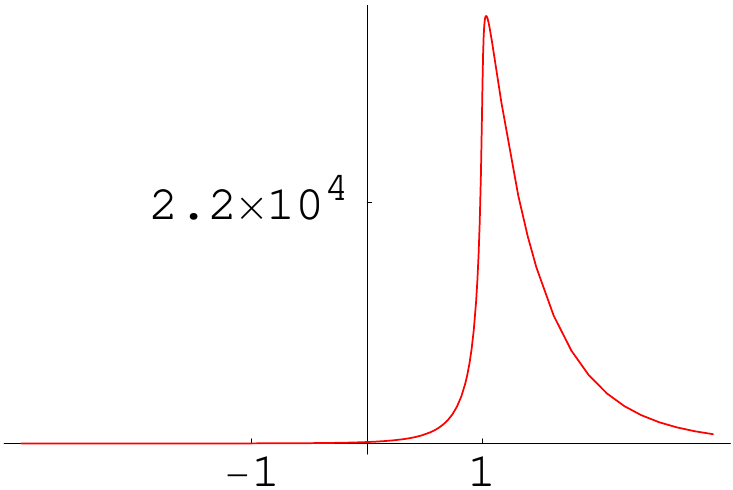}}  \bsep{2ex}
\\
 \hline \tsep{2ex} $\begin{array}{@{}c@{}} {\rm Fermionic}\\ {\rm zero\  mode}\end{array}$ &
\parbox{4cm}{\includegraphics[width=4cm]{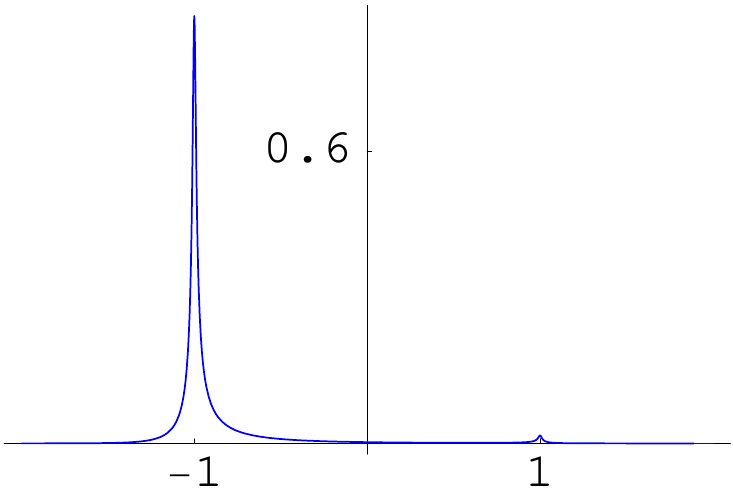}} &
\parbox{4cm}{\includegraphics[width=4cm]{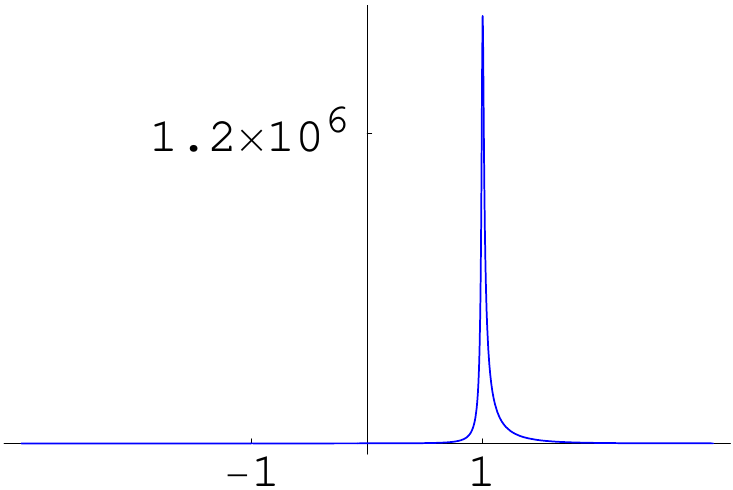}} &
\parbox{4cm}{\includegraphics[width=4cm]{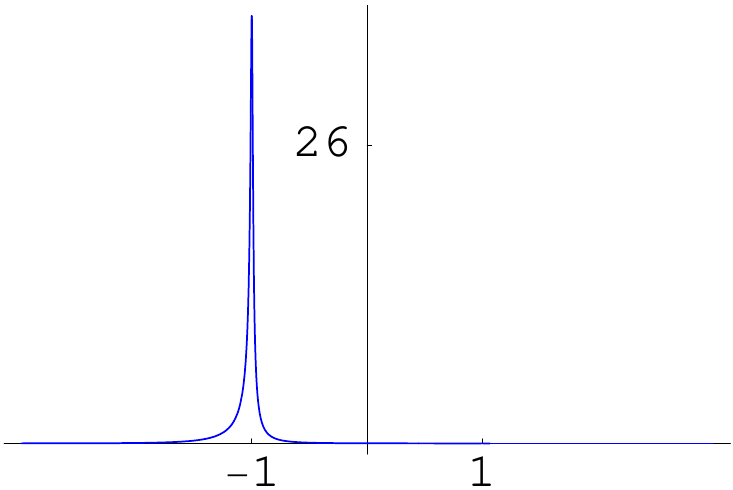}}  \bsep{2ex} \\
 \hline
 \end{tabular}
\end{table}

\begin{table}[htdp]\centering
\caption{$\bar{\hbar} = 2$, $\delta=1/2$.}

\vspace{1mm}

\begin{tabular}{|@{}c@{}|@{}c@{\,\,}c@{\,\,}c@{}|} \hline  \tsep{2ex}
$|\Psi_{0}^{(0/1)}(x_1,x_2)|^2$ &  Type I  &  $\begin{array}{@{}c@{}} {\rm Type \ IIa}\\ \kappa=3\end{array}$
  &  $\begin{array}{@{}c@{}} {\rm Type \ IIb} \\ \kappa=3 \end{array}$ \bsep{1ex} \\
 \hline \tsep{2ex} $\begin{array}{@{}c@{}} {\rm Bosonic}\\ {\rm zero \ mode}\end{array}$   &
\parbox{4cm}{\includegraphics[width=4cm]{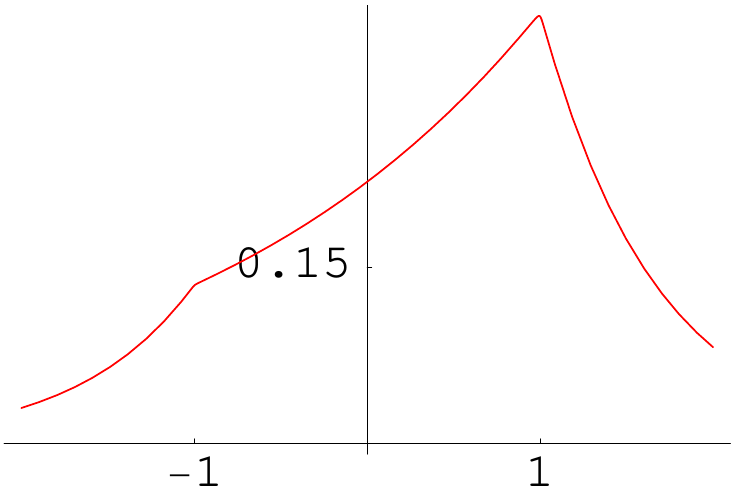}} &
\parbox{4cm}{\includegraphics[width=4cm]{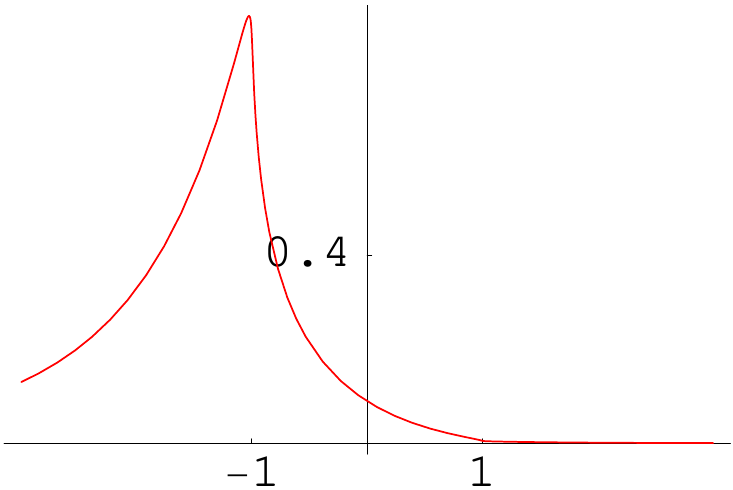}} &
\parbox{4cm}{\includegraphics[width=4cm]{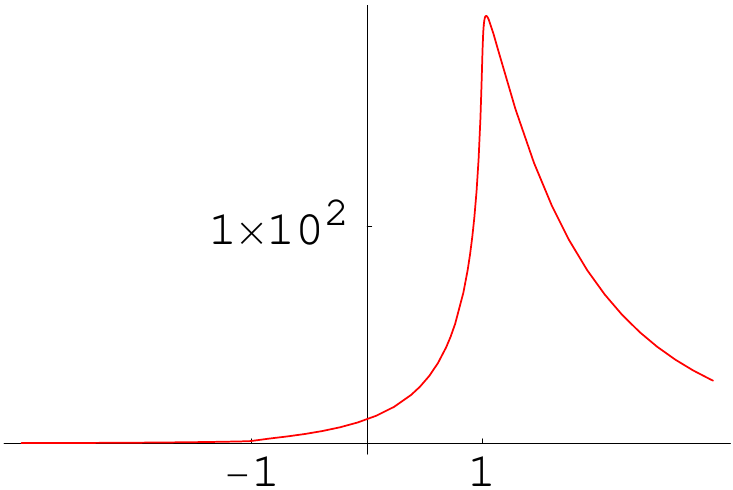}}  \bsep{2ex}
\\
 \hline \tsep{2ex} $\begin{array}{@{}c@{}} {\rm Fermionic}\\ {\rm zero \ mode}\end{array}$ &
\parbox{4cm}{\includegraphics[width=4cm]{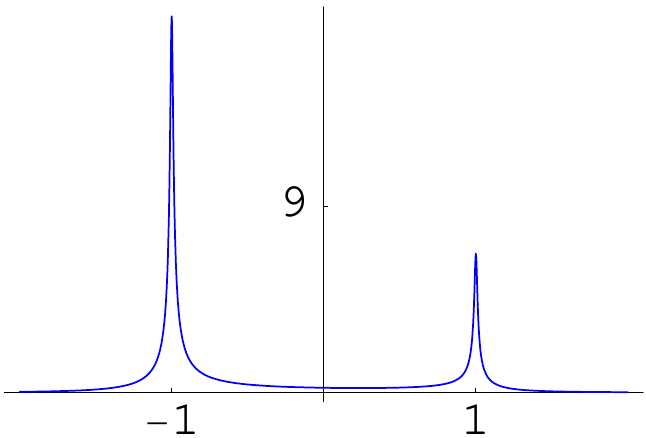}} &
\parbox{4cm}{\includegraphics[width=4cm]{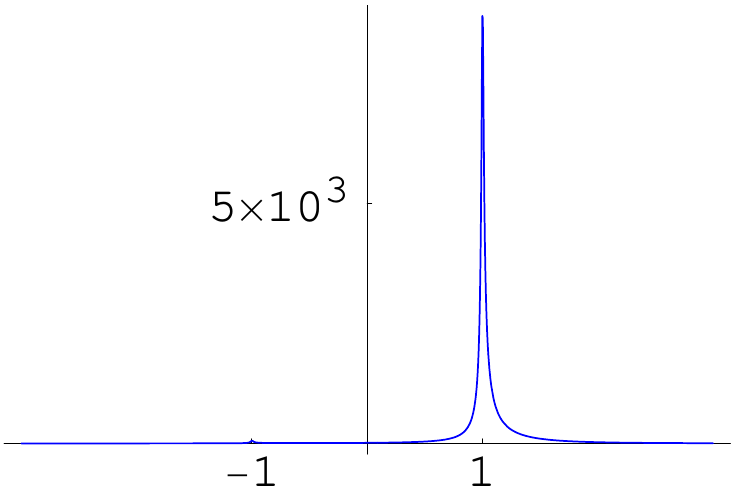}} &
\parbox{4cm}{\includegraphics[width=4cm]{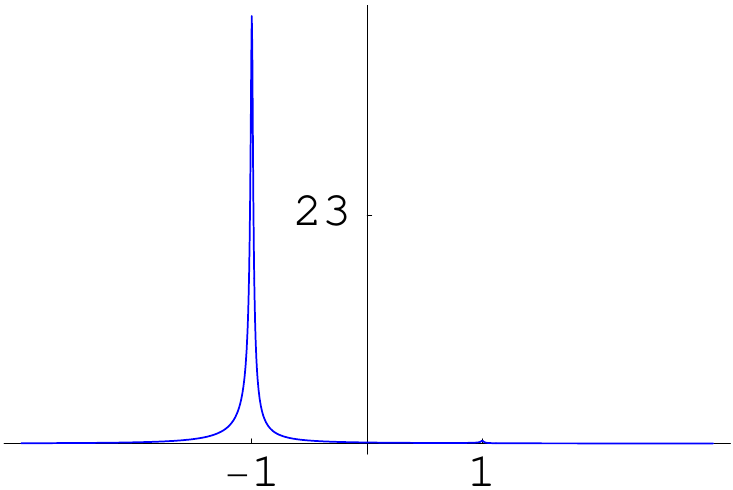}}  \bsep{2ex} \\
 \hline
 \end{tabular}
\end{table}

\begin{table}[htdp]\centering
\caption{$\bar{\hbar} = 4$, $\delta=1/2$.}
\vspace{1mm}

\begin{tabular}{|@{}c@{}|@{}c@{\,\,}c@{\,\,}c@{}|}  \hline  \tsep{2ex}
$|\Psi_{0}^{(0/1)}(x_1,x_2)|^2$ &  Type I  &  $\begin{array}{@{}c@{}} {\rm Type \ IIa}\\ \kappa=3\end{array}$
  &  $\begin{array}{@{}c@{}} {\rm Type \ IIb} \\ \kappa=3 \end{array}$ \bsep{1ex} \\
 \hline \tsep{2ex} $\begin{array}{@{}c@{}} {\rm Bosonic}\\ {\rm zero \ mode}\end{array}$   &
\parbox{4cm}{\includegraphics[width=4cm]{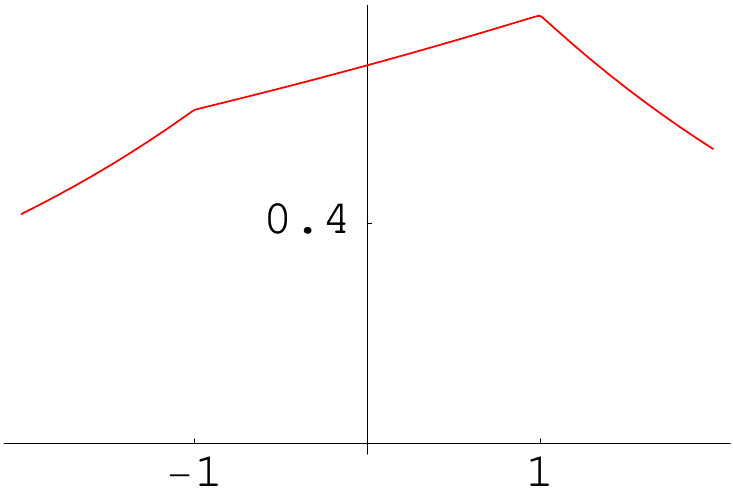}} &
\parbox{4cm}{\includegraphics[width=4cm]{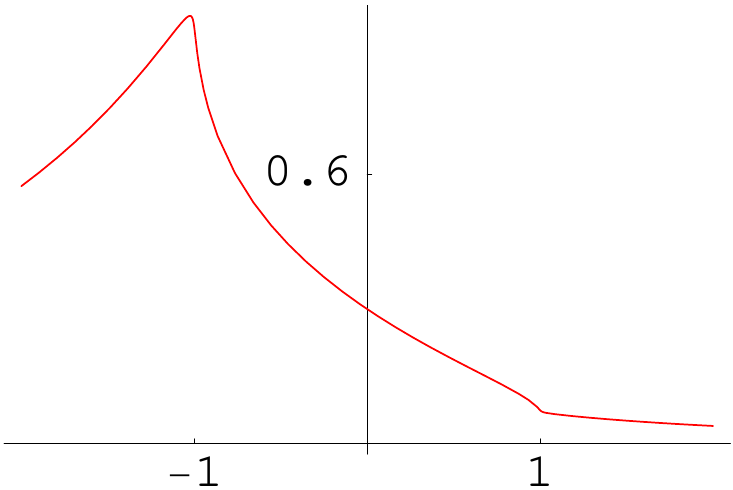}} &
\parbox{4cm}{\includegraphics[width=4cm]{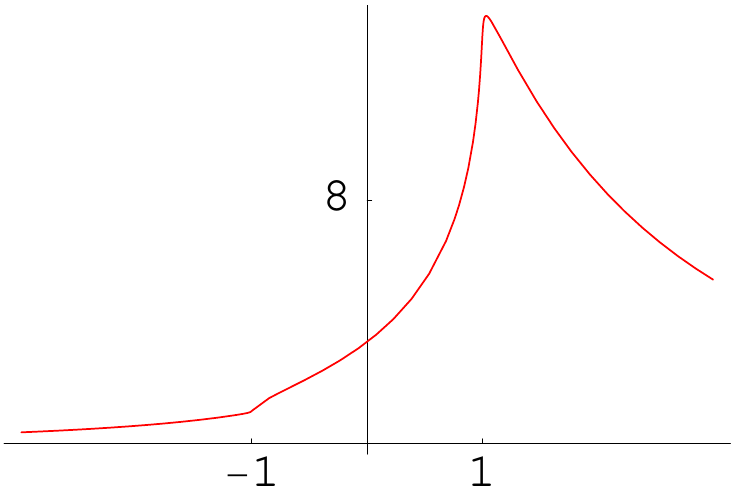}}  \bsep{2ex}
\\
 \hline \tsep{2ex} $\begin{array}{@{}c@{}} {\rm Fermionic}\\ {\rm zero \ mode}\end{array}$ &
\parbox{4cm}{\includegraphics[width=4cm]{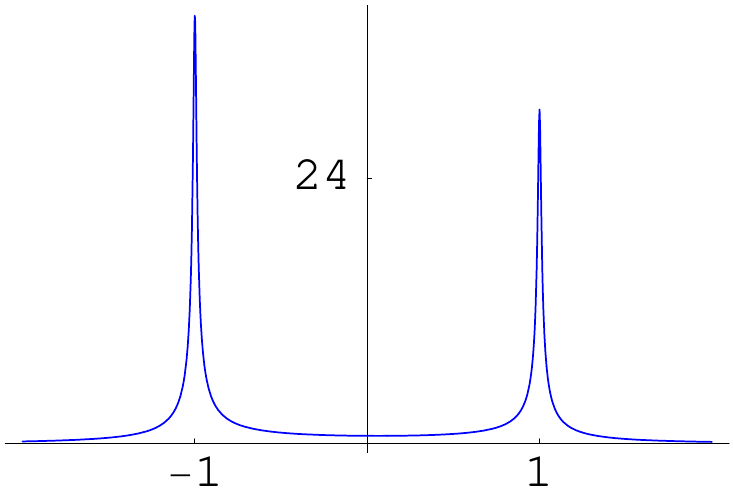}} &
\parbox{4cm}{\includegraphics[width=4cm]{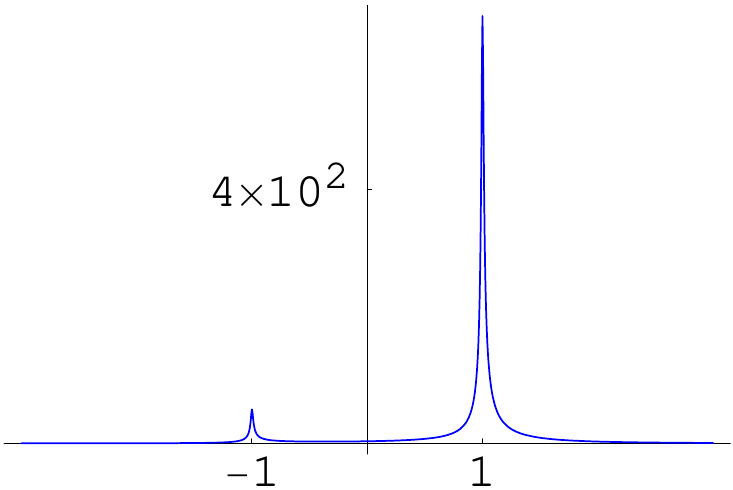}} &
\parbox{4cm}{\includegraphics[width=4cm]{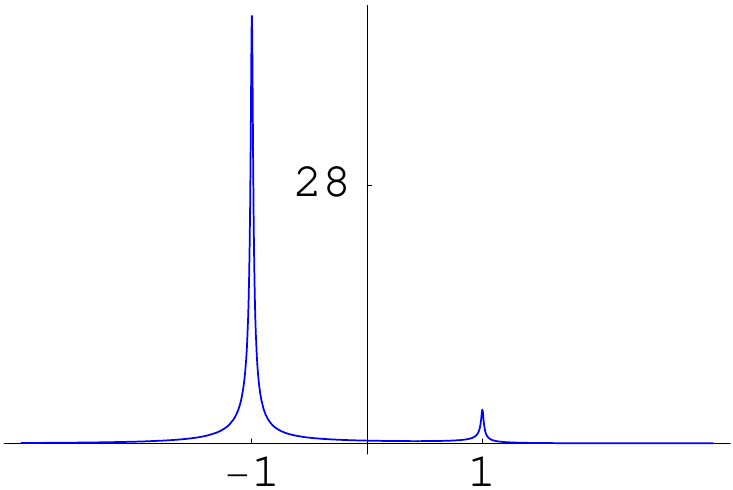}} \bsep{2ex} \\
 \hline
 \end{tabular}
\end{table}

\subsection*{Acknowledgements}

We are grateful to Mikhail Iof\/fe for many enlightening lessons and
conversations on SUSY quantum mechanics as well as letting us know
about reference \cite{Bondar}. JMG thanks Nigel Hitchin for sending
him his unpublished lecture notes on the Dirac operator.

Finally, we recognize f\/inancial support from the Spanish DGICYT and
the Junta de Castilla y Le\'on under contracts: FIS2006-09417,
VAO13C05.

\pdfbookmark[1]{References}{ref}
\LastPageEnding

\end{document}